  \documentclass[
  aps,
  prd,
  twocolumn,
  showpacs,
  superscriptaddress
  ]{revtex4-2}

  \usepackage[english]{babel}
  \usepackage[dvips]{color}
  \usepackage{natbib}
  \usepackage{latexsym}
  \usepackage{epsfig}
  \usepackage{amsmath}
  \usepackage{amssymb}
  \usepackage{multirow}
  \usepackage{tabularx}
  \usepackage{longtable}
  \usepackage{wasysym}

  \def\etal{\emph{et\,al.}}
  \def\ndf{\text{ndf}}
  \def\NRB{\text{NRB}}

  \def\CCmnA{\nu_{\mu}p\to\mu^-p\pi^+}
  \def\CCmnB{\nu_{\mu}n\to\mu^-p\pi^0}
  \def\CCmnC{\nu_{\mu}n\to\mu^-n\pi^+}
  \def\CCmnD{\nu_{\mu}p\to\mu^-\Delta^{++}}

  \def\CCmaA{\overline{\nu}_{\mu}n\to\mu^+n\pi^-}
  \def\CCmaB{\overline{\nu}_{\mu}p\to\mu^+n\pi^0}
  \def\CCmaC{\overline{\nu}_{\mu}p\to\mu^+p\pi^-}

  \newcommand{\MR}[2]{\multirow{#1}*{#2}}
  \newcommand{\MC}[3]{\multicolumn{#1}{#2}{#3}}

\begin{document}

  \newcolumntype{C}{>{\centering}X}
  \newcolumntype{Y}{>{\centering\arraybackslash}X}
  \newcolumntype{K}{>{\centering\arraybackslash\arraybackslash}X}
  \providecommand{\rs}{{\fontencoding{U}\fontfamily{pzd}\selectfont\Red{\symbol{51}}}}

  \title{Resonance axial-vector mass from experiments on \\
         neutrino-hydrogen and neutrino-deuterium scattering}

  \author{Igor~D.~Kakorin}
  \email{Kakorin@jinr.ru}
  \affiliation{N.~N.~Bogoliubov Laboratory of Theoretical Physics, \\
               Joint Institute for Nuclear Research, RU-141980 Dubna, Russia}

  \author{Konstantin~S.~Kuzmin}
  \email{KKuzmin@theor.jinr.ru}
  \affiliation{N.~N.~Bogoliubov Laboratory of Theoretical Physics, \\
               Joint Institute for Nuclear Research, RU-141980 Dubna, Russia}
  \affiliation{A.~I.~Alikhanov Institute for Theoretical and Experimental Physics \\
               of NRC ``Kurchatov Institute'', RU-117218 Moscow, Russia}

  \date{\today} 

  \begin{abstract}
  We analyze all available experimental data on the $\nu_\mu$ and $\overline{\nu}_\mu$
  total and differential cross sections of charged-current single pion production
  through the decay of intermediate nucleon and baryon resonances
  measured on hydrogen and deuterium targets
  in the accelerator experiments at ANL, BNL, FNAL, and CERN.
  These data are used to determine
  the current ``resonance'' axial-vector mass of the nucleon
  and to fine tune the nonresonance noninterfering background contribution
  which described within the Rein-Sehgal approach.
  For this analysis,
  we revise the phenomenological model and
  the experimental dataset for the fits,
  modify the method of likelihood analysis
  compared to the previous study.
  The obtained model parameters coming in combination with 
  a revised strategy for the normalization of the Breit-Wigner distributions are
  slightly different from the values used by default.
  \end{abstract}



  \maketitle

  \section{Introduction}


  A precise calculation of the neutrino-nucleon 
  scattering cross sections is extremely important 
  for a correct interpretation of the results 
  in studying neutrino properties
  in atmospheric and accelerator experiments
  \cite{FernandezMartinez:2010dm,%
        Meloni:2012fq,%
        Benhar:2013oba,%
        Coloma:2013rqa,%
        Coloma:2013tba,%
        Jen:2014aja,%
        Ericson:2015cva,%
        Ankowski:2016bji,%
        Ankowski:2016jdd,%
        Mosel:2019vhx}.
  The cross sections from charged current quasielastic scattering (CC QES),
  1$\pi$-production reactions through the decay of baryon and nucleon resonances,
  and deep inelastic scattering
  are crucial for the few-GeV neutrino experiments
  since all the contributions are comparable 
  in the neutrino energy range of about 1~GeV \cite{Kuzmin:2005bm,Kuzmin:2006dt}.

  For the phenomenological description of the resonance 1$\pi$-production reactions,
  we use the phenomenological Ravndal's model \cite{Ravndal:1973xx},
  which was revised and modified by Rein and Sehgal in 1981 (RS model) \cite{Rein:1980wg,Rein:1987cb}, 
  and other authors in subsequent years.  
  The RS model is based on the
  relativistic harmonic-oscillator quark model
  in the formulation by Feynman, Kislinger, and Ravndal (FKR) \cite{Feynman:1971wr}
  taking into account contributions from
  eighteen interfering low-lying nucleon and baryon resonances
  with masses $\lesssim 2$~GeV
  and a noninterfering nonresonance contribution as a background (NRB) 
  for resonance reactions.
  The FKR approach has been adopted for the RS model
  in the assumption of the standard dipole parametrization
  for the vector and axial-vector transition form factors,
  \begin{equation*}
  G^{V,A}\left(Q^2\right)\propto\left(1+\frac{Q^2}{4M_N^2}\right)^{1/2-n}
  \left(1+\frac{Q^2}{M_{V,A}^2}\right)^{-2},
  \end{equation*}
  with the ``standard'' value of the vector mass $M_V= 0.84$~GeV and
  $n$--number of excitation in the resonance.
  In the original version of the RS model
  the resonance axial mass was chosen as $M_A= 0.95$~GeV,
  which is the average result of
  ANL 12-foot bubble chamber \cite{Derrick:1978jz} and
  CERN Gargamelle \cite{Rollier:1978kr,Dewit:1978} experiments
  to study QES $\nu_\mu$ and $\overline{\nu}_\mu$ reactions.
  In our previous study \cite{Kuzmin:2006dh},
  the value of $M_A$ was obtained from the global likelihood analysis
  of all available experimental data known at that time
  from the measurement of the total cross sections
  of the 1$\pi$-production reactions with a variety of nuclear targets.
  The world average value of $M_A= 1.12 \pm 0.03$ GeV
  obtained in that analysis is used in several Monte Carlo neutrino generators
  as a default or an option
  \cite{Andreopoulos:2009rq,%
        Andreopoulos:2015wxa,%
        McGivern:2016bwh,%
        Adamson:2014pgc,%
        Abe:2011ks}.
  In the present paper, we propose a new global fit of $M_A$
  based on the clarified phenomenological model,
  improved method of likelihood analysis, and
  the revised experimental dataset.
  Let us recall that the parameter $M_A$ is purely phenomenological and
  differs for different reaction types and different models of 
  the electromagnetic form factors of the nucleon and
  the parametrization of the axial-vector form factor.

  \begin{figure*}[htb]
  \includegraphics[width=0.97\linewidth]{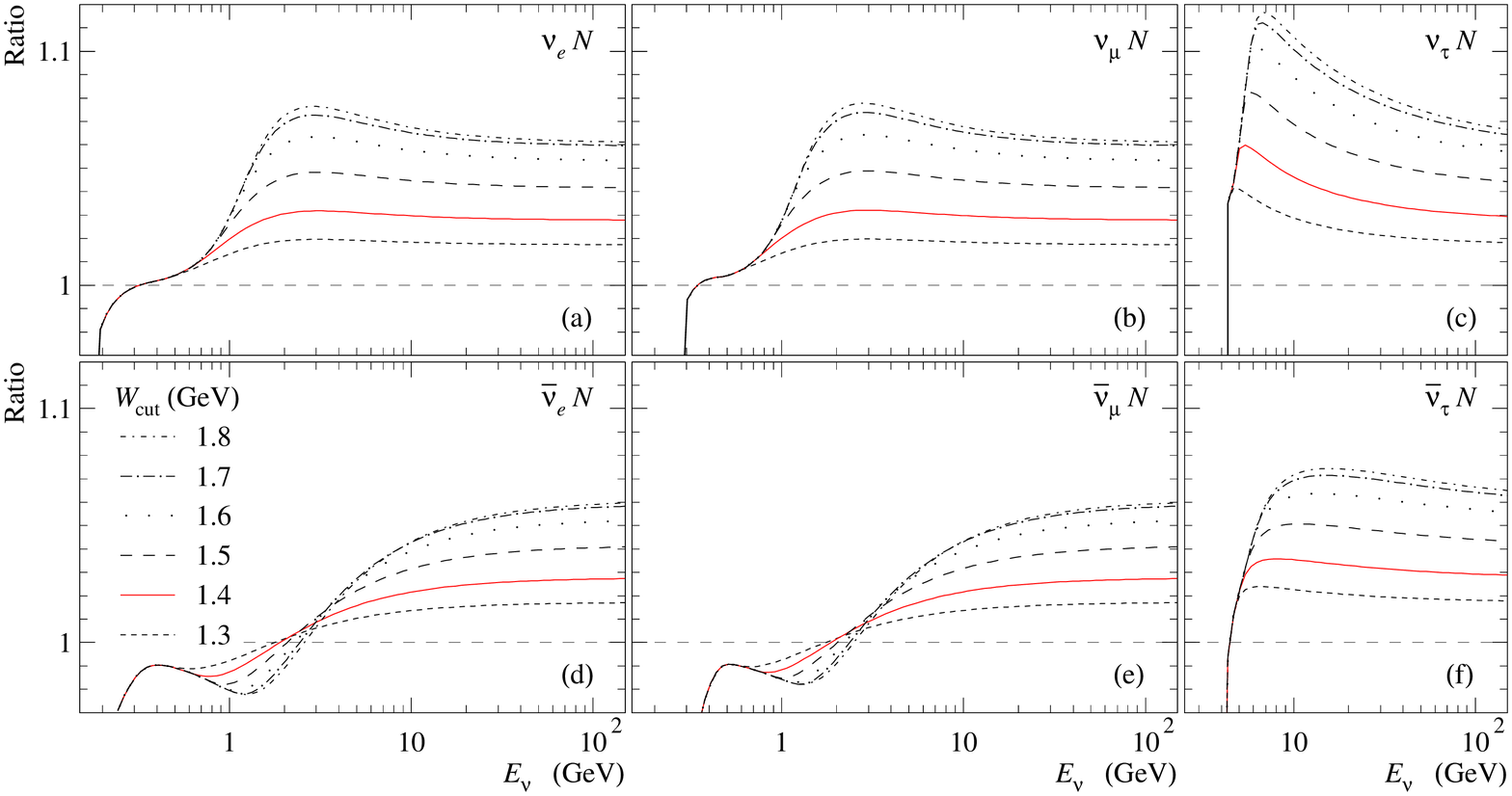}
  \caption{(Color online)
           Ratios of the total cross sections for the resonance 1$\pi$-production reactions
           predicted with $M_A$ obtained from the global fits for H$_2$ and D data
           without and with the resonance interference effect
           as a function of neutrino energy.
           In order to only illustrate the interference total effect,
           the cross sections are calculated without NRB contributions
		   for six values of $W$ from $1.3$ GeV to $1.8$ GeV
           for the reactions of
           $\nu_e$ (a),
           $\nu_\mu$ (b),
           $\overline{\nu}_\mu$ (c),
           $\overline{\nu}_e$ (d),
           $\overline{\nu}_\mu$ (e), and
           $\overline{\nu}_\tau$ (f)
		   with free isoscalar nucleons.
          }
  \label{Fig:rRESCC_interference_test_2_BSc}
  \end{figure*}


  The original version of the RS model
  neglects the mass of the final charged lepton.
  This feature of the model is not a significant disadvantage
  for the description of
  the $\nu_e$, $\overline{\nu}_e$, $\nu_\mu$, and $\overline{\nu}_\mu$ reactions
  but is a crucial gap
  for the description of the $\nu_\tau$ and $\overline{\nu}_\tau$ reactions
  (it should be noted that Rein and Sehgal proposed the model at the end of the 1970s,
  and at that time the possibility of experimentally detecting
  the $\nu_\tau$ or $\overline{\nu}_\tau$ reaction was not discussed).
  In 2004, a model was adopted
  to calculate the cross sections
  of the $\nu_\tau$ and $\overline{\nu}_\tau$ by using the covariant form
  of the charged leptonic current with a definite lepton helicity,
  keeping the hadronic current unchanged
  (the so-called extended RS model or KLN model) \cite{Kuzmin:2003ji,Kuzmin:2004ya}.
  Inclusion of the final charged lepton mass
  leads to a slight decreasing of the cross sections in the case of light leptons
  and to a notable one in the case of the $\tau$-lepton.
  In 2006, Berger and Sehgal
  took into account the pion-pole contribution
  to the hadronic axial current (KLN-BS model) \cite{Berger:2007rq}.
  The pion-pole contribution reduces the cross sections by a few percent
  in the neutrino energy range at the threshold of the reactions
  and decreases rapidly with increasing neutrino energy.
  In 2008, Graczyk and Sobczyk
  investigated different approaches
  to accounting for the nonzero mass of the charged lepton
  and the modification of the axial current due to the pion pole-term.
  It was shown that their result is equivalent
  to the predictions of the KLN-BS model.
  Alternative vector and axial form factors are proposed
  to improve the RS approach in the
  $\Delta(1232)$ resonance region \cite{Graczyk:2008zz,Graczyk:2007bc}.
  In 2018, Kabirnezhad \cite{Kabirnezhad:2017xzx}
  following the original paper by Rein \cite{Rein:1987cb}
  and ideas of Hern$\acute{\text{a}}$ndez, Nieves, and Valverde \cite{Hernandez:2007qq},
  suggested a new approach for calculating the interfering NRB
  for the RS model \cite{Kabirnezhad:2017jmf,Kabirnezhad:2016nwu,Kabirnezhad:2017dui}.
  This modification of the Rein model contains 17 interfering resonances with masses below 2~GeV
  [the $F_{17}(1900)$ is excluded from the model
  because evidence of its existence appears only in PDG 2020 \cite{Zyla:2020zbs}
  and it does not have a significant effect on calculations).

  Rein and Sehgal represented NRB by a resonance amplitude of $P_{11}$ character (like the nucleon)
  with the Breit-Wigner factor replaced by an adjustable constant $f_\NRB$.
  The corresponding cross section of NRB reactions
  is added incoherently to the cross sections of resonance productions,
  assuming that NRB contributions are smooth in
  the phase-spacelike kinematic region and affect only the cross section
  of final states with $I = 1/2$.
  Ignoring subtle effects of coherence NRB is regulated by $f_\NRB$ only.
  The problem is that the value of $f_\NRB$ cannot be predicted theoretically in the RS approach.
  Probably, the condition of $f_\NRB = 1$ was chosen by the authors of the model
  to optimize the description of selected experimental data relevant at that time.
  Therefore, $f_\NRB$ as well as $M_A$
  should be defined from the modern global fits of experimental data.

  The noninterfering NRB
  is determined by one more free phenomenological parameter $W_\NRB$,
  which is the bound for the invariant mass of the final hadron system, $W$,
  separating the kinematic region of the resonance 1$\pi$-production
  and deep inelastic reactions.
  In other words, the parameter $W_\NRB$ controls
  the applicability range of NRB contributions in the RS model.
  The parameter $W_\NRB$, as well as $f_\NRB$,
  is not based on the theory and is fixed in the RS model 
  by default to a sufficiently large asymptotic value of 2.5 GeV.
  The reasonable value of $W_\NRB$ is important 
  for describing experimental data measured without cuts for $W$.
  Furthermore, the dependence of the cross sections on the parameter $W_\NRB$
  is significant for neutrino energies above $\sim 10$ GeV.
  There are no reliable experimental data
  for the cross sections of the resonance 1$\pi$-production reactions
  measured without cut for $W$ in the neutrino energy range
  above a few tens of GeV.
  So the parameter $W_\NRB$ cannot be reliably obtained
  from the global fit.
  For all the fits the value of $W_\NRB$ is chosen to be the default value.

  Theoretical uncertainties of predicted NRB reactions
  and model-dependent nuclear effects in neutrino-nucleus interactions
  complicate the physical interpretation of the phenomenological parameter $M_A$
  obtained from the global fit of the experimental data
  measured on a variety of nuclear targets.
  To avoid this complication, we use the experimental data
  measured only on hydrogen and deuterium targets.
  In the RS approach,
  the $\CCmnD$, $\CCmnA$, and $\CCmaA$ reactions
  do not require NRB a for description of the experimental data
  in contrast with all other resonance 1$\pi$-production reactions.
  We suggest using $M_A$ obtained from the global fit
  for the cross sections of the $\nu_\mu$ and $\overline{\nu}_\mu$ reactions
  not requiring NRB.
  Using the world average value of $M_A$,
  we found the adjustable constant $f_\NRB$
  from the global fit with a fixed value of $M_A$
  for the experimental data on the cross sections of the reactions requiring NRB.


  The version of the KLN-BS model implemented into the
  GENIE Monte Carlo neutrino event generator currently 
  does not take into account the interference of the amplitudes of resonances
  with the same spin and orbital-angular momentum.
  Neutrino interaction with nuclei 
  leads to the generation of hadron resonances with different quantum numbers.
  The amplitudes of resonances have to interfere to produce the 
  final state of the hadron system under consideration.
  Each of the interfering resonances, by simultaneous decay,
  produces the same final system
  with one or several pions at a fixed invariant final mass $W$.
  Figure \ref{Fig:rRESCC_interference_test_2_BSc} illustrates the effect of interference
  for the total cross sections of the reactions on isoscalar nucleons.
  The interference effect is highly reliant on the final hadron-system mass and
  is equal to a few per cent for all neutrino energies.
  It should be noted that the cross section calculated with and without taking into account 
  the interference of resonances may require somewhat different values of axial mass
  for describing the same experimental dataset.

  In this study, all the cross sections 
  are calculated with original software package
  taking into account of interference and using the
  physical constants published in PDG 2020 \cite{Zyla:2020zbs},
  whereas the official physical tunes of all GENIE versions
  are obtained with earlier values of the parameters
  according to PDG 2016 \cite{Patrignani:2016xqp}.
  Differences between the current and previous values of
  the central mass, total width, and the branching ratio of resonance reactions
  with a single pion in the final state
  have a negligible effect on the cross sections and results of the global fits.


  \section{Normalizations of the Breit-Wigner distributions}

\begin{figure}[htb!]
  \includegraphics[width=0.97\linewidth]{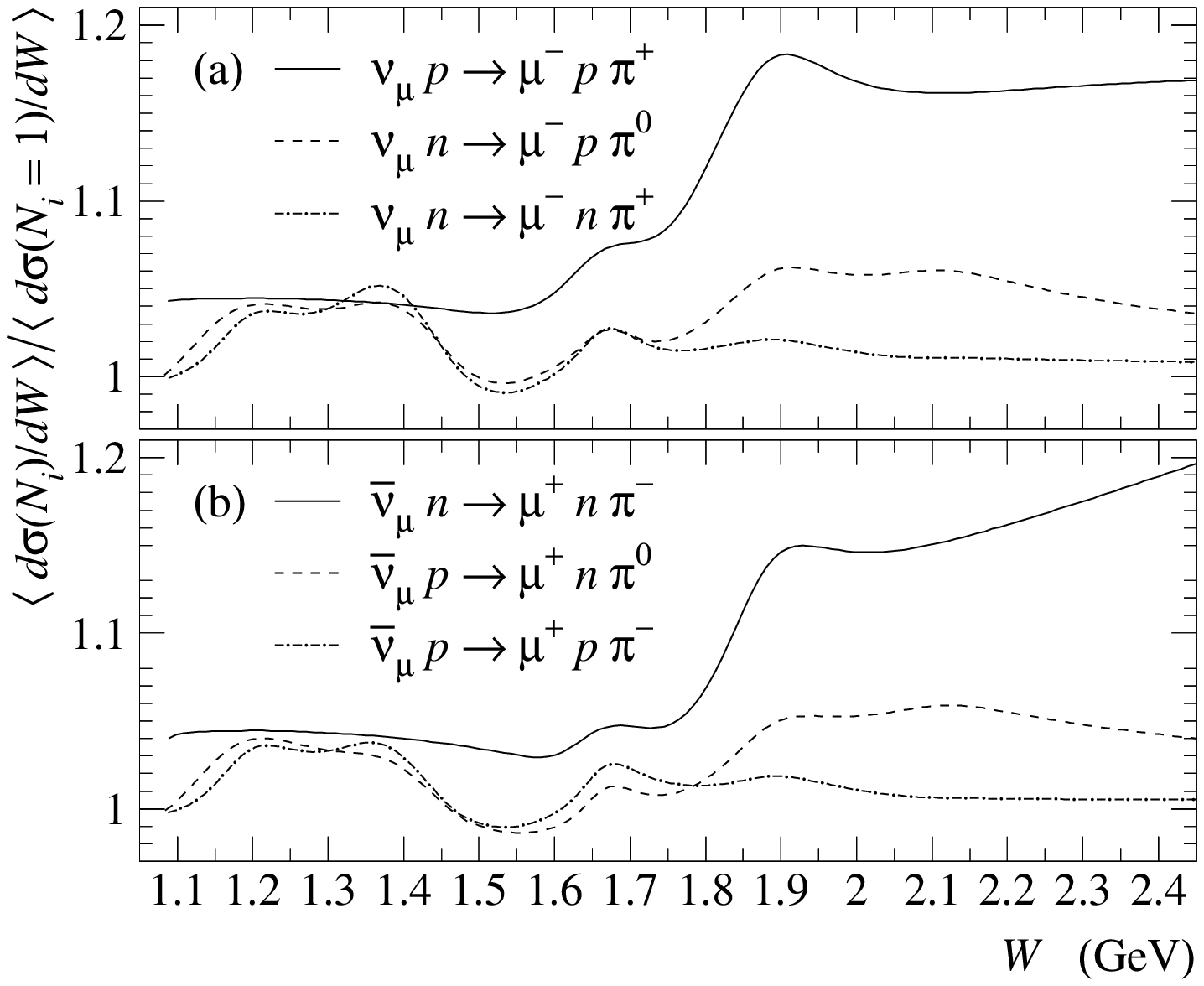}
  \caption{
           Ratios of the differential averaged overs the BNL wide-band
           $\nu_{\mu}(\overline{\nu}_\mu)$ beams \cite{Ahrens:1986ke}
           cross sections as functions of $W$
           of the reactions $\CCmnA$, $\CCmnB$, $\CCmnC$ (a) and
           $\CCmaA$, $\CCmaB$, $\CCmaC$ (b)
           calculated with normalizations of Breit-Wigner factors
           used in the previous version of KLN-BS model
           (listed in the notation to the text)
           to the corresponding cross sections
           predicted with the definition of $N_i= 1$.
           The cross sections are shown for
           $M_A$ and $f_\NRB$ obtained from the global fit.
          }  
  \label{Fig:dsRESCC_dW_N_Breit-Wigner_test}
  \end{figure}

 
  In all the previous versions of the model (RS, KLN, and KLN-BS)
  the $W$-dependent Breit-Wigner distributions of resonances
  are normalized to correction factors, $N_i$, which are not well defined.
  We discuss this problem below and suggest how to avoid them.
  The normalization factors of the Breit-Wigner distributions,
  $\eta^i_{BW}\left(W\right)$, are defined by Eq.~(2.33) in \cite{Rein:1980wg}

  \begin{align*}
                                N_i = &\, \int^\infty_{W_{\text{min}}}\tilde{\eta}^i_{BW}\left(W\right) dW, \\
  \tilde{\eta}^i_{BW}\left(W\right) = &\, \frac{1}{2\pi}\dfrac{\Gamma_i\left(W\right)}{\left(W-M^2_i\right)^2+\Gamma^2_i\left(W\right)/4}, \\
             \Gamma_i\left(W\right) = &\, \Gamma^0_i\left[\dfrac{q_\pi\left(W\right)}{q_\pi\left(M_i\right)}\right]^{2L+1}, \\
                q_\pi\left(W\right) = &\, \dfrac{\sqrt{\left(W^2-m^2_{N'}-m^2_\pi\right)^2-4m^2_{N'}m^2_\pi}}{2W},
  \end{align*}
  where
  $W_{\text{min}} = m_{N'}+m_\pi$,
  $m_{N'}$, $m_\pi$, $L$, $M_i$, and $\Gamma^0_i$ are
  the masses of the final nucleon and pion,
  total orbital-angular momentum of resonance,
  resonance mass,
  and the Breit-Wigner width, respectively.
  To calculate $N_i$ numerically,
  we need to define the cutoff for the upper limit of $W$.
  The result of integration strongly depends on the cutoff value
  but there are no physical reasons to choose a definite value for it.
  \begin{figure}[htb]
  \includegraphics[width=0.97\linewidth]{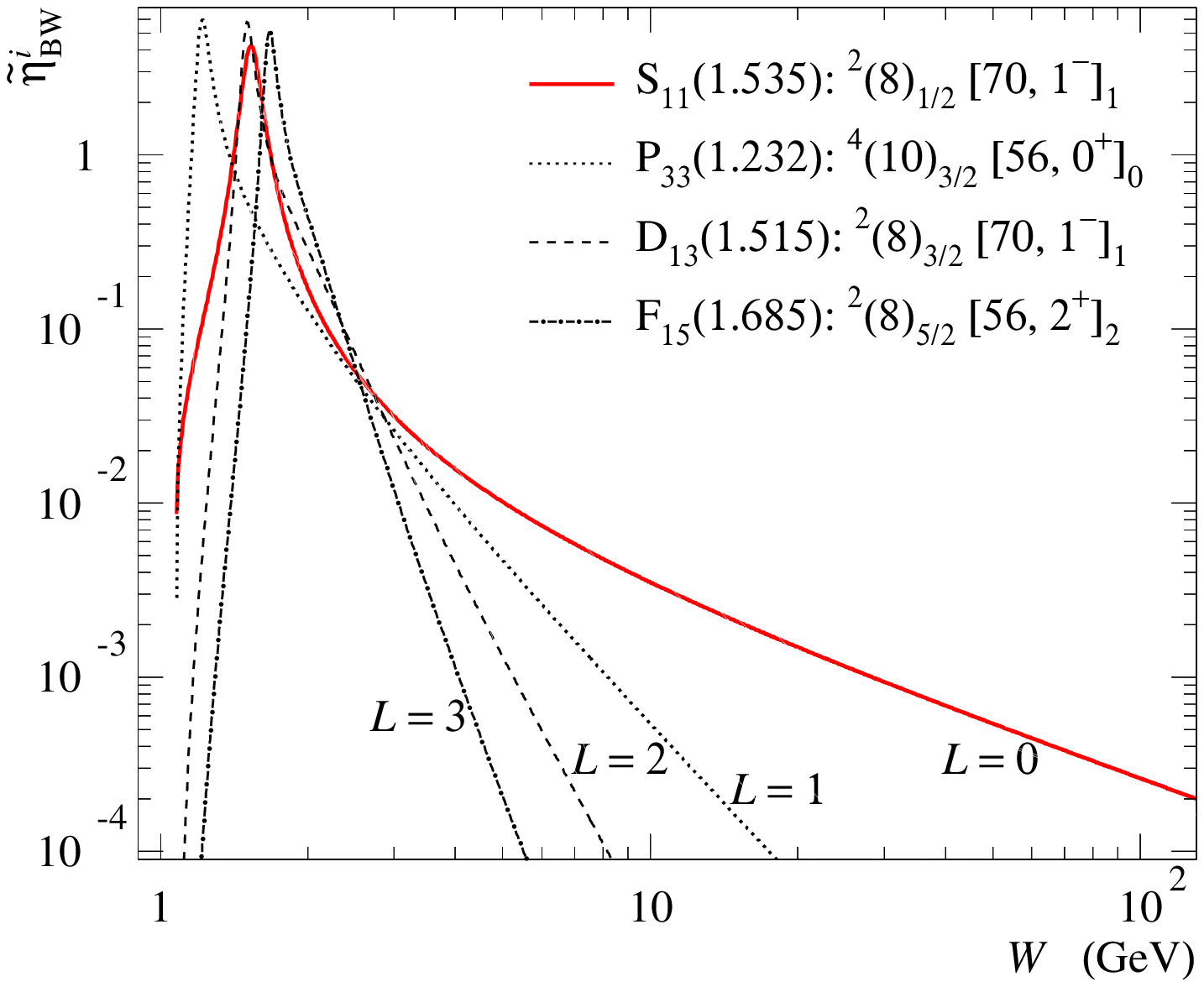}
  \caption{(Color online)
           Breit-Wigner distributions, $\tilde{\eta}^i_{BW}\left(W\right)$,
           as functions of the invariant mass of the final hadron system, $W$,
           for the lightest
           $S_{11}$ ($L=0$),
           $P_{33}$ ($L=1$),
           $D_{13}$ ($L=2$), and
           $F_{15}$ ($L=3$)
           resonance states in the $\CCmnB$ reaction.
           The solid line indicates that the $\sim 1/W$ function
           leads to unphysical $N_S=\infty$.
           }
  \label{Fig:Breit-Wigner}
  \end{figure}
  Figure \ref{Fig:Breit-Wigner} shows a typical dependence
  of $\tilde{\eta}^i_{BW}\left(W\right)$ on the variable $W$
  for the $S$, $P$, $D$, and $F$ resonance states.
  Asymptotic behavior of the $S$ resonances is $\sim 1/W$
  unlike the $P$, $D$, and $F$ resonances.
  The trivial change of the variable $W = W_{\text{min}}/W'$ leads to
  the limits of integration from 0 to 1 for the expression of $N_{i=S}$.
  The integrand becomes $\tilde{\eta}^i_{BW}\left(W'\right) \sim 1/W'$,
  and its integration gives the nonphysical value for $N_S=\infty$.
  Rein abandoned using the normalization factors
  of the Breit-Wigner distributions in \cite{Rein:1987cb}.
  The definition of $N_i = 1$ is the simplest possibility
  to avoid the ambiguity in the calculation.
  Figure \ref{Fig:dsRESCC_dW_N_Breit-Wigner_test}
  shows the ratios of the differential flux-weighted cross sections,
  $\langle{d\sigma/dW}\rangle$,
  for the $\nu_\mu$ and $\overline{\nu}_\mu$ resonance reactions 
  with the values of $N_i$
  \footnote{
  The normalizations of the resonance Breit-Wigner distributions
  according to the previous version of KLN-BS model are
  0.957 for $P_{33}(1234)$,
  0.784 for $P_{11}(1450)$,
  1.055 for $S_{31}(1620)$,
  0.935 for $P_{33}(1640)$,
  0.751 for $D_{33}(1730)$,
  1.229 for $P_{31}(1920)$,
  0.635 for $F_{35}(1920)$,
  0.710 for $F_{37}(1950)$,
  1.285 for $P_{33}(1960)$,
  1.008 for $D_{13}(1525)$,
  1.067 for $S_{11}(1540)$,
  1.051 for $S_{11}(1640)$,
  1.165 for $D_{13}(1670)$,
  1.024 for $D_{15}(1680)$,
  0.912 for $F_{15}(1680)$,
  1.349 for $P_{11}(1710)$,
  1.301 for $P_{13}(1740)$, and
  0.619 for $F_{17}(1970)$.}
  used in the previous version of KLN-BS model
  to the corresponding cross sections predicted without normalizations.
  The cross sections are averaged over the energy spectra 
  from the BNL wide-band $\nu_{\mu}$ and $\overline{\nu}_\mu$
  beams up to 7 GeV \cite{Ahrens:1986ke}.
  At this energy range the kinematical region of $W$
  covers whole the mass range of resonances included in the RS-based models.
  The greatest effect from the normalizations is achieved at hight values of $W$
  and does not significantly affect the total cross sections.
  The total cross sections calculated with the clear definition of $N_i$
  are a few percent lower in comparison with
  ones calculated with ambiguously determined values of normalizations.

  The GENIE proposes the {\em ad hoc} modification of the Breit-Wigner distributions,
  $\Gamma'\left(W\right)$, for the $\Delta$ resonance
  \begin{align*}
  \Gamma'\left(W\right)= &\,
  \Gamma_{N\pi}\Gamma\left(W\right)+ 
  \Gamma_{N\gamma}\Gamma^0
  \left[\dfrac{1+c/q_\gamma\left(M\right)}
              {1+c/q_\gamma\left(W\right)}\right]^2
        \dfrac{q_\gamma\left(W\right)}{q_\gamma\left(M\right)},\\
  q_\gamma\left(W\right) = &\, \dfrac{W^2-m^2_{N'}}{2W},
  \end{align*}
  where the function $\Gamma\left(W\right)$ is defined by the above equation for $L= 1$,
  $\Gamma_{N\pi}$, and $\Gamma_{N\gamma}$ are the decay-branching
  ratios of resonance to $N\pi$ and $N\gamma$ states,
  $M$ is the central value of the mass of the $P_{33}(1232)$ resonance,
  $c= 0.706$ GeV.
  The modified $\Gamma'\left(W\right)$ function is very close to
  the original $\Gamma\left(W\right)$ function
  because the value of $\Gamma_{N\pi}$ is close to 1 (0.994)
  and the value of $\Gamma_{N\gamma}$ is close to zero (0.006).
  In the present calculations
  we do not use the GENIE modification of the Breit-Wigner function.

  \section{Method of data fitting}

  We use the following least-square statistical model
  \begin{equation}
  \label{Blobel_fixed}
  \chi^2 = \sum_{i}\left\{\sum_{j \in G_i}\frac{\left[N_iT_{ij}(\boldsymbol{\lambda})-E_{ij}\right]^2}{\sigma_{ij}^2}
           +\frac{\left(N_i-1\right)^2}{\sigma_i^2}\right\}.
  \end{equation}
  In this equation,
  index $i$ enumerates the experimental data groups $G_i$, 
  index $j \in G_i$ enumerates the bin-averaged experimental data $E_{ij}$ from the group $G_i$, and
  $\sigma_{ij}$ is the error of $E_{ij}$,
  without the uncertainty due to the $\nu/\overline{\nu}$ flux normalization.
  The flux normalization $N_i$, individual for each data group $G_i$ 
  is treated as a free-fitting parameter and is included into the ordinary penalty term,
  $\left(N_i-1\right)^2/\sigma_i^2$, where $\sigma_i$ is the flux normalization error.
  The $T_{ij}(\boldsymbol{\lambda})$ represents a bin-averaged theoretical prediction
  depending on a set of fitting parameters $\boldsymbol{\lambda}$.
  The procedure of minimization can be simplified by substituting 
  $N_i=\mathcal{N}_i(\boldsymbol{\lambda})$,
  into Eq.~\eqref{Blobel_fixed}
  where $\mathcal{N}_i(\boldsymbol{\lambda})$
  are obtained from the analytic solution to the equations $\partial\chi^2/\partial N_i=0$,
  \begin{equation*}
  \label{Eqn:Normalization_Factor}
  \mathcal{N}_i(\boldsymbol{\lambda})
  = \dfrac{1+\sigma_i^2\sum_{j \in G_i}\sigma_{ij}^{-2}T_{ij}(\boldsymbol{\lambda})E_{ij}}
          {1+\sigma_i^2\sum_{j \in G_i}\sigma_{ij}^{-2}T_{ij}^2(\boldsymbol{\lambda})}.
  \end{equation*}
  As it follows from the analysis, the deviation of the normalization factors $N_i$
  from unity for each data group $G_i$ does not exceed
  the doubled experimental uncertainty of the corresponding
  $\nu_\mu$ or $\overline{\nu}_\mu$ flux normalization.
 
  All the fits are done with the CERN function minimization
  and error analysis package MINUIT (version 94.1) \cite{James:1994vla,James:1975343},
  thus taking care of getting an accurate error matrix.
  The errors of the output parameters quoted below correspond to
  the usual one and two-standard deviation ($1\sigma$ and $2\sigma$) errors.
  Numerical calculations with GENIE are done with
  the JINR cloud infrastructure of Multifunctional Information and Computing Complex
  \cite{Baranov:2016gvt,*Balashov:2018}.

  \section{Experimental dataset}

  We analyzed and classified all available 1$\pi$-production experimental data, 
  presented as the $\nu_\mu$  and $\overline{\nu}_\mu$ scattering cross sections  
  on hydrogen and deuterium targets measured in
  ANL \cite{Campbell:1973wg,%
            Schreiner:1973ka,%
            Derrick:1974ema,%
            Perkins:1975bj,%
            Barish:1975bw,%
            Barish:1977qk,%
            Barnes:1978cs,%
            Barish:1978pj,%
            Derrick:1980nr,%
            Derrick:1980xw,%
            Radecky:1981fn},
  BNL \cite{Lee:1976wr,%
            Fanourakis:1980si,%
            Baker:1980pj,%
            Baker:1983qc,%
            Kitagaki:1986ct},  
  FNAL \cite{Ross:1978yu,%
             Bell:1978rb,Bell:1978qu,%
             Barish:1979ny}, and
  CERN \cite{Franzinetti:276222,%
             Young:1967ud,%
             Budagov:1969pw,%
             Hasert:1975sv,%
             Schmid:1978yt,%
             Krenz:1977sw,%
             Lerche:1978cp,%
             Pohl:1978pi,Pohl:1979fw,%
             Bolognese:1978yz,Bolognese:1979gf,%
             Allen:1980ti,%
             Allasia:1983qh,%
             Barlag:1984uga,%
             Allen:1985ti,%
             Jones:1989vt,%
             Allasia:1983dq,%
             Allasia:1990uy}
  accelerator experiments
  from the beginning of the 1970s to the 1990s
  with energy ranges from about hundreds of MeV (ANL) to tens of GeV (CERN).

  {\squeezetable
  \begin{table*}[!htb]
  \caption{
           Experimental datasets and
           data tapes involved into the global fit of $M_A$,
           results of individual fits for $M_A$ of selected datasets,
           normalization factors $N_{\nu,\overline{\nu}}$,
           absolute and normalized to $\ndf$ values of $\chi^2$
           (only absolute value in the case of $\ndf= 0$),
           where $\ndf=\text{NP}-N_p$,
           NP is the number of experimental data bins in the dataset, and
           $N_p= 1, 2$ is the number of free parameters.
           Statistical errors of the fitted parameters
           correspond to one and two (shown in brackets) standard deviations.
           The datasets of
           ANL 1982 \cite{Radecky:1981fn} and
           BNL 1986 \cite{Kitagaki:1986ct}
           include the original data and the data recalculated by
           Furuno~\etal\, \cite{Furuno:2003ng-proc,Sakuda:2002-KK,Sakuda:2003-KK} and
           Rodrigues~\etal\, \cite{Rodrigues:2016xjj}.
          }
  \center{ 
  \begin{tabularx}{\linewidth}{lccCr}                                                                                                                                                                                                                                 \hline\hline\noalign{\smallskip}
  \label{Tab:MA_RES_individual_fits_D_H_2_no_NRB}
   Experiment                                                    & Data                                                        &\MC{1}{c}{$M_A$\,(GeV)}               & $N_{\nu,\overline{\nu}}$             &\MC{1}{c}{$\chi^2/\ndf$}  \\ \noalign{\smallskip}       \hline\noalign{\smallskip}
   Barish~\etal, ANL 1979 \cite{Barish:1978pj}                   &$\langle{d\sigma\left(\CCmnD\right)/dQ^2}\rangle$            &$1.03_{-0.13\,(0.20)}^{+0.15\,(0.26)}$&$1.09_{-0.13\,(0.22)}^{+0.14\,(0.22)}$& $7.27/\phantom{0}7=1.04$ \\ \noalign{\smallskip}
   Radecky~\etal, ANL 1982 \cite{Radecky:1981fn}                 &$\sigma\left(\CCmnA\right)$                                  &$1.16_{-0.06\,(0.10)}^{+0.07\,(0.11)}$&$1.04_{-0.04\,(0.07)}^{+0.04\,(0.07)}$&$90.19/          21=4.29$ \\ \noalign{\smallskip}
                                                                 &$\sigma\left(\CCmnA\right)/\sigma^{\text{QES}}$              &$1.13_{-0.09\,(0.17)}^{+0.09\,(0.18)}$&$1$                                   &$12.93/\phantom{0}6=2.16$ \\ \noalign{\smallskip}
   Kitagaki~\etal, BNL 1986 \cite{Kitagaki:1986ct}               &$\sigma\left(\CCmnD\right)\&\,\sigma\left(\CCmnA\right)$     &$1.20_{-0.12\,(0.18)}^{+0.14\,(0.24)}$&$1.07_{-0.11\,(0.18)}^{+0.12\,(0.20)}$&$ 9.53/          13=0.73$ \\ \noalign{\smallskip}
                                                                 &$\sigma\left(\CCmnA\right)/\sigma^{\text{QES}}$              &$1.25_{-0.03\,(0.06)}^{+0.03\,(0.06)}$&$1$                                   &$23.60/           8=2.95$ \\ \noalign{\smallskip}
   Bell~\etal, FNAL 1978 \cite{Bell:1978rb,Bell:1978qu}          &$\sigma\left(\CCmnA\right)$                                  &$1.11_{-0.18\,(0.29)}^{+0.20\,(0.33)}$&$1.00_{-0.16\,(0.26)}^{+0.16\,(0.26)}$&$ 0.02$                   \\ \noalign{\smallskip}
   Allen~\etal, CERN BEBC 1980 \cite{Allen:1980ti}               &$\sigma\left(\CCmnD\right)$                                  &$1.09_{-0.13\,(0.20)}^{+0.15\,(0.25)}$&$1.00_{-0.15\,(0.24)}^{+0.15\,(0.24)}$&$ 6.54/\phantom{0}5=1.31$ \\ \noalign{\smallskip}
   Allasia~\etal, CERN BEBC 1983 \cite{Allasia:1983qh}           &$\sigma\left(\CCmaA\right)$                                  &$0.98_{-0.15\,(0.24)}^{+0.17\,(0.28)}$&$1.00_{-0.15\,(0.25)}^{+0.15\,(0.25)}$&$ 2.24/\phantom{0}4=0.56$ \\ \noalign{\smallskip}
   Barlag~\etal, CERN BEBC 1984 \cite{Barlag:1984uga}            &$\sigma\left(\CCmnA\right)$                                  &$1.10_{-0.14\,(0.22)}^{+0.16\,(0.27)}$&$1.00_{-0.15\,(0.24)}^{+0.15\,(0.25)}$&$ 1.36/\phantom{0}4=0.34$ \\ \noalign{\smallskip}
   Allen~\etal, CERN BEBC 1986 \cite{Allen:1985ti}               &$\sigma\left(\CCmnA\right)$, $W < 2$ GeV                     &$1.07_{-0.13\,(0.20)}^{+0.16\,(0.26)}$&$1.00_{-0.15\,(0.24)}^{+0.15\,(0.24)}$&$ 4.38/\phantom{0}3=1.46$ \\ \noalign{\smallskip}
                                                                 &$\sigma\left(\CCmnA\right)$, $W > 2$ GeV                     &$0.98_{-0.19\,(0.31)}^{+0.19\,(0.33)}$&$1.00_{-0.15\,(0.25)}^{+0.15\,(0.25)}$&$ 3.84/\phantom{0}2=1.92$ \\ \noalign{\smallskip}
   Allasia~\etal, CERN BEBC 1990 \cite{Allasia:1990uy}           &$\sigma\left(\CCmnA\right)$                                  &$1.15_{-0.11\,(0.18)}^{+0.13\,(0.22)}$&$1.00_{-0.13\,(0.21)}^{+0.13\,(0.21)}$&$ 3.69/\phantom{0}4=0.92$ \\ \noalign{\smallskip}
                                                                 &$\sigma\left(\CCmaA\right)$                                  &$1.29_{-0.12\,(0.18)}^{+0.14\,(0.23)}$&$1.00_{-0.13\,(0.21)}^{+0.13\,(0.21)}$&$ 2.29/\phantom{0}4=0.57$ \\ \noalign{\smallskip}\hline\hline\noalign{\smallskip}
  \end{tabularx}}
  \end{table*}
  }

  {\squeezetable
  \begin{table*}[!htb]
  \caption{
           The same items as in Table \ref{Tab:MA_RES_individual_fits_D_H_2_no_NRB}
           but for the datasets involved into the individual and
           global fits of $f_\NRB$.
           The data recalculated by
           Furuno~\etal\, \cite{Furuno:2003ng-proc,Sakuda:2002-KK,Sakuda:2003-KK} and
           Rodrigues~\etal\, \cite{Rodrigues:2016xjj}
           marked by indices $^{\text{F}}$ and $^{\text{R}}$, respectively.
          }
  \label{Tab:FB_RES_individual_fits_D_H_2_with_NRB}
  \center{ 
  \begin{tabularx}{\linewidth}{lcCCr}                                                                                                                                                                                                                                                               \hline\hline\noalign{\smallskip}
   Experiment                                         & Data                                                    & $f_\NRB$                                     & $N_{\nu,\overline{\nu}}$                     &\MC{1}{c}{$\chi^2/\ndf$}                    \\ \noalign{\smallskip}\hline      \noalign{\smallskip}
   Radecky~\etal, ANL 1982 \cite{Radecky:1981fn}      &$\sigma\left(\CCmnB\right)\&\,\sigma\left(\CCmnC\right)$,&\MR{2}{$1.24_{-0.22\,(0.38)}^{+0.19\,(0.31)}$}&\MR{2}{$1.00_{-0.05\,(0.09)}^{+0.05\,(0.09)}$}&\MR{2}{          $16.20/\phantom{0}6=2.70$} \\
                                                      &$W < 1.6$ GeV                                            &                                              &                                              &                                            \\ \noalign{\smallskip}
                                                      &$\sigma\left(\CCmnB\right)\&\,\sigma\left(\CCmnC\right)$,&\MR{2}{$1.40_{-0.14\,(0.24)}^{+0.13\,(0.21)}$}&\MR{2}{$1.00_{-0.04\,(0.06)}^{+0.04\,(0.06)}$}&\MR{2}{          $69.48/          22=3.16$} \\
                                                      &$W < 1.4$ GeV, no cut$^{\text{R}}$                       &                                              &                                              &                                            \\ \noalign{\smallskip}
                                                      &$\sigma\left(\CCmnB\right)/\sigma^{\text{QES}}$,         &\MR{2}{$1.21_{-0.46\,( -  )}^{+0.33\,(0.59)}$}&\MR{2}{ 1}                                    &\MR{2}{$\phantom{0}9.59/\phantom{0}3=3.20$} \\
                                                      &no cut$^{\text{F}}$                                      &                                              &                                              &                                            \\ \noalign{\smallskip}
                                                      &$\sigma\left(\CCmnC\right)/\sigma^{\text{QES}}$,         &\MR{2}{$1.14_{-0.19\,(0.43)}^{+0.17\,(0.31)}$}&\MR{2}{ 1}                                    &\MR{2}{          $13.03/\phantom{0}4=3.26$} \\
                                                      &no cut$^{\text{F}}$                                      &                                              &                                              &                                                \\ \noalign{\smallskip} 
   Kitagaki~\etal, BNL 1986 \cite{Kitagaki:1986ct}    &$\sigma\left(\CCmnB\right)$,                             &\MR{2}{$1.29_{-0.42\,(0.74)}^{+0.42\,(0.71)}$}&\MR{2}{$1.10_{-0.19\,(0.30)}^{+0.19\,(0.31)}$}&\MR{2}{          $28.17/\phantom{0}8=3.52$} \\
                                                      &no cut                                                   &                                              &                                              &                                            \\ \noalign{\smallskip}
                                                      &$\sigma\left(\CCmnC\right)$,                             &\MR{2}{$1.00_{-0.27\,(0.45)}^{+0.30\,(0.52)}$}&\MR{2}{$0.99_{-0.20\,(0.32)}^{+0.20\,(0.32)}$}&\MR{2}{          $44.81/\phantom{0}9=4.98$} \\
                                                      &no cut                                                   &                                              &                                              &                                            \\ \noalign{\smallskip}
                                                      &$\sigma\left(\CCmnB\right)/\sigma^{\text{QES}}$,         &\MR{2}{$1.61_{-0.10\,(0.19)}^{+0.09\,(0.17)}$}&\MR{2}{ 1}                                    &\MR{2}{          $25.82/\phantom{0}8=3.23$} \\
                                                      &no cut$^{\text{F}}$                                      &                                              &                                              &                                            \\ \noalign{\smallskip}
                                                      &$\sigma\left(\CCmnC\right)/\sigma^{\text{QES}}$,         &\MR{2}{$0.91_{-0.08\,(0.16)}^{+0.07\,(0.13)}$}&\MR{2}{ 1}                                    &\MR{2}{          $58.03/\phantom{0}8=7.25$} \\
                                                      &no cut$^{\text{F}}$                                      &                                              &                                              &                                            \\  \noalign{\smallskip}
   Allasia~\etal, CERN BEBC 1983 \cite{Allasia:1983qh}&$\sigma\left(\CCmaC\right)$,                             &\MR{2}{$1.18_{-0.31\,(0.54)}^{+0.29\,(0.48)}$}&\MR{2}{$1.00_{-0.15\,(0.25)}^{+0.15\,(0.24)}$}&\MR{2}{$\phantom{0}6.53/\phantom{0}4=1.63$} \\
                                                      &$W < 2$ GeV                                              &                                              &                                              &                                            \\ \noalign{\smallskip}
   Barlag~\etal, CERN BEBC 1984 \cite{Barlag:1984uga} &$\sigma\left(\CCmnB\right)\&\,\sigma\left(\CCmnC\right)$,&\MR{2}{$0.72_{-0.21\,(0.37)}^{+0.19\,(0.30)}$}&\MR{2}{$0.92_{-0.10\,(0.16)}^{+0.10\,(0.16)}$}&\MR{2}{          $33.01/          10=3.30$} \\
                                                      &$W < 2$ GeV                                              &                                              &                                              &                                            \\ \noalign{\smallskip}
   Allen~\etal, CERN BEBC 1986 \cite{Allen:1985ti}    &$\sigma\left(\CCmaC\right)$,                             &\MR{2}{$0.25\, \pm 100\%$}                    &\MR{2}{$1.00_{-0.15\,(0.25)}^{+0.12\,(0.18)}$}&\MR{2}{$\phantom{0}0.86/\phantom{0}3=0.29$} \\
                                                      &$W < 2$ GeV                                              &                                              &                                              &                                            \\ \noalign{\smallskip}
                                                      &$\sigma\left(\CCmaC\right)$,                             &\MR{2}{$0.88_{-0.19\,(0.32)}^{+0.18\,(0.30)}$}&\MR{2}{$1.01_{-0.15\,(0.23)}^{+0.15\,(0.23)}$}&\MR{2}{$\phantom{0}2.45/\phantom{0}2=1.22$} \\
                                                      &$W > 2$ GeV                                              &                                              &                                              &                                            \\ \noalign{\smallskip}
   Allasia~\etal, CERN BEBC 1990 \cite{Allasia:1990uy}&$\sigma\left(\CCmaC\right)$,                             &\MR{2}{$1.30_{-0.20\,(0.31)}^{+0.22\,(0.38)}$}&\MR{2}{$1.00_{-0.15\,(0.25)}^{+0.15\,(0.25)}$}&\MR{2}{          $10.84/\phantom{0}4=2.71$} \\
                                                      &$W < 2$ GeV                                              &                                              &                                              &                                            \\ \noalign{\smallskip}
                                                      &$\sigma\left(\CCmnC\right)+\sigma\left(\CCmaC\right)$,   &\MR{2}{$1.10_{-0.14\,(0.36)}^{+0.23\,(0.37)}$}&       $0.85$                                 &\MR{2}{          $27.12/\phantom{0}3=9.04$} \\
                                                      &$W > 2$ GeV                                              &                                              &       $0.99$                                 &                                            \\ \noalign{\smallskip}\hline\hline\noalign{\smallskip}
  \end{tabularx}}
  \end{table*}
  }

  \begin{figure*}[htb!]
  \includegraphics[width=0.97\linewidth]{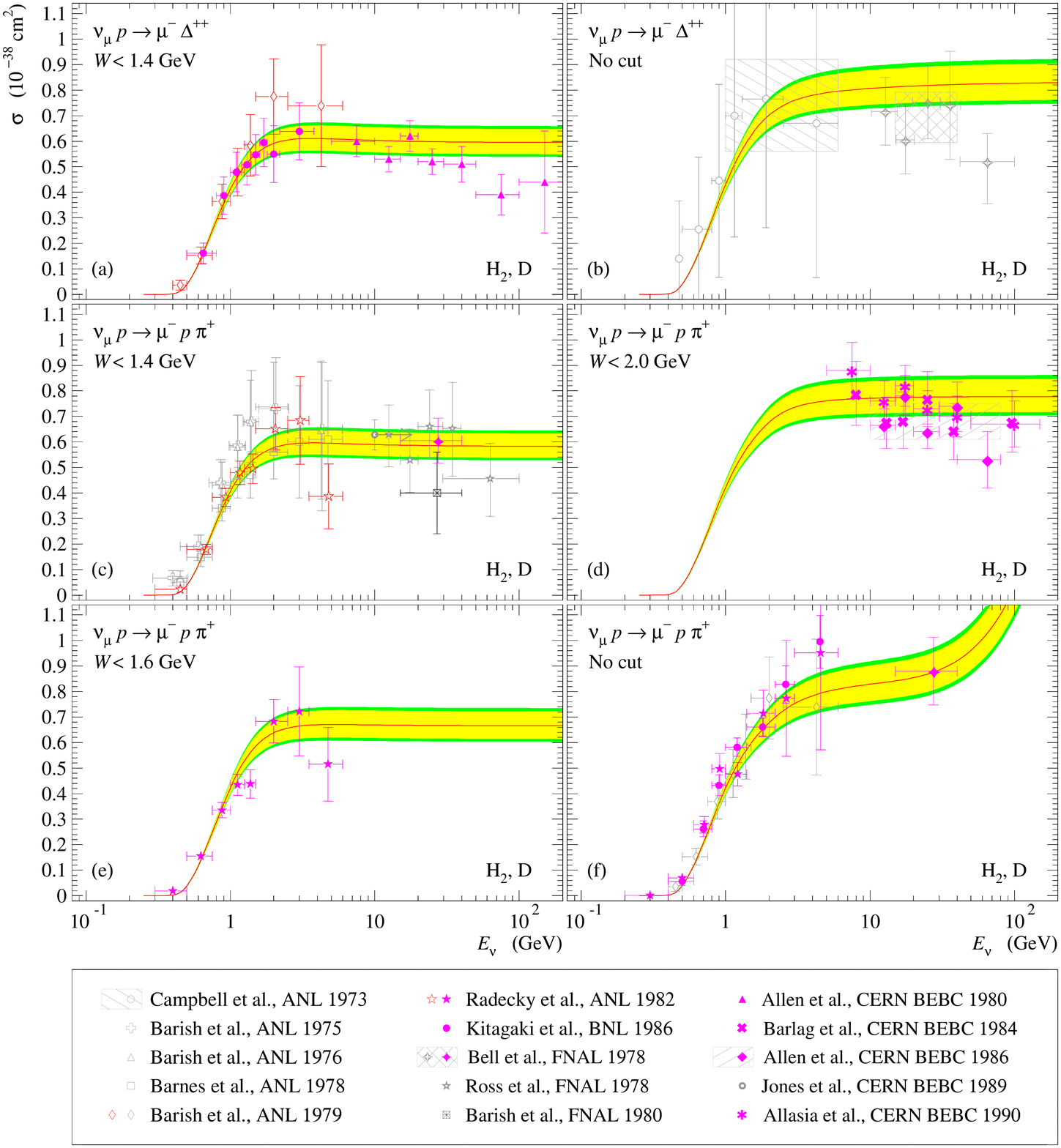}
  \caption{(Color online)
           Total cross sections of the reactions $\CCmnD$ (a), (b) and
           $\CCmnA$ (c) -- (f) as functions of neutrino energy
           predicted with $M_A$ obtained from the global fit for H$_2$ and D data
           in comparison with the experimental data measured at
           ANL 1973 \cite{Campbell:1973wg},
               1975 \cite{Perkins:1975bj},
               1976 \cite{Barish:1975bw},
               1978 \cite{Barnes:1978cs},
               1979 \cite{Barish:1978pj},
               1982 \cite{Radecky:1981fn},
           BNL 1986 \cite{Kitagaki:1986ct},
           FNAL 1978 \cite{Bell:1978rb,Ross:1978yu},
                1980 \cite{Barish:1979ny}, and
           CERN BEBC 1980 \cite{Allen:1980ti},
                     1984 \cite{Barlag:1984uga},
                     1986 \cite{Allen:1985ti},
                     1989 \cite{Jones:1989vt},
                     1990 \cite{Allasia:1990uy}.
           The original data of
           ANL 1982 \cite{Radecky:1981fn} and
           BNL 1986 \cite{Kitagaki:1986ct} are
           recalculated by Rodrigues~\etal\, \cite{Rodrigues:2016xjj}.
           Error bars of the experimental data points show
           the statistical and systematical errors added quadratically
           with no errors of $\nu_\mu$ flux normalizations.
           Inner and outer shaded bands around the curves demonstrate
           the uncertainty of the fitted value of $M_A$
           corresponding to $1\sigma$ and $2\sigma$ errors, respectively.
           Titles of the reactions, targets, and experimental cuts on $W$
           are given in the legends in each panel.
           }
  \label{Fig:sRESCC_D_withoutNRB_103.1.31.301.h4_2_BSc}
  \end{figure*}

  \begin{figure*}[htb!]
  \includegraphics[width=0.97\linewidth]{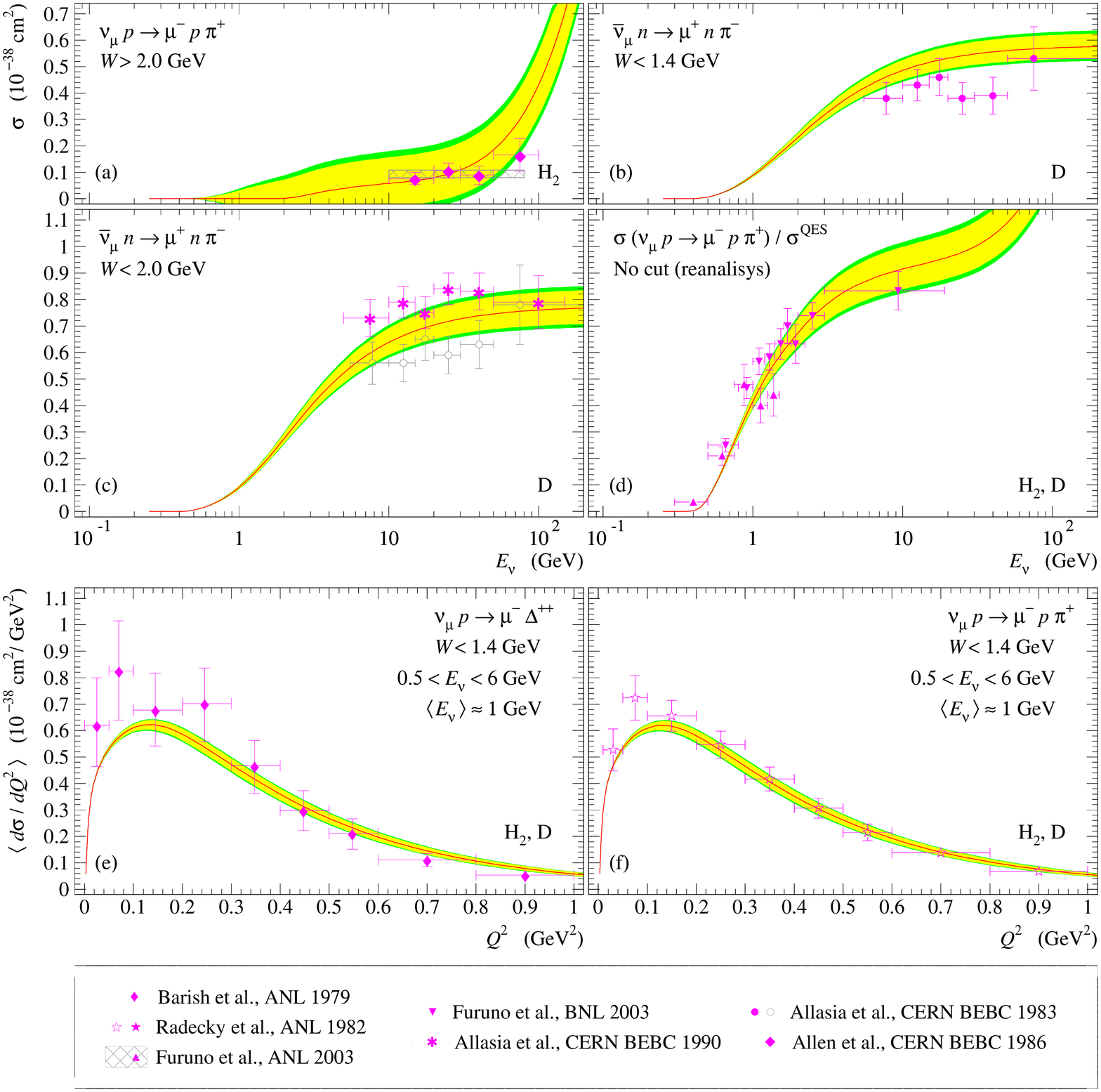}
  \caption{(Color online)
           Total cross sections 
           of the reaction $\CCmnA$ (a), $\CCmaA$ (b), (c),
           ratio of the total cross sections of the reaction $\CCmnA$ to
           the quasielastic reaction $\nu_{\mu}n\to\mu^-p$ (d)
           as functions of neutrino energy and
           $Q^2$-dependent flux-weighted differential cross sections
           of the reactions $\CCmnD$ (e) and $\CCmnA$ (f)
           predicted with $M_A$ obtained from the global fit for H$_2$ and D data
           in comparison with the experimental data measured at
           ANL 1979 \cite{Barish:1978pj},
               1982 \cite{Radecky:1981fn} and
           CERN BEBC 1983 \cite{Allasia:1983qh},
                     1986 \cite{Allen:1985ti},
                     1990 \cite{Allasia:1990uy}.
           Differential cross sections are averaged over the flux
           borrowed from \cite{Barish:1977qk}.
           Titles of the reactions, targets, 
           experimental cuts on $W$, and
           additional information are given in the legends.
           Notation for the solid curves and shaded bands is the same as in
           Fig. \ref{Fig:sRESCC_D_withoutNRB_103.1.31.301.h4_2_BSc}.
          }
  \label{Fig:sRESCC+dsRESCC_dQ2_D_withoutNRB_103.1.31.301.h4_2_BSc_101.3.31.301.2k_2_BBBA25}
  \end{figure*}

  \begin{figure*}[htb!]
  \includegraphics[width=0.97\linewidth]{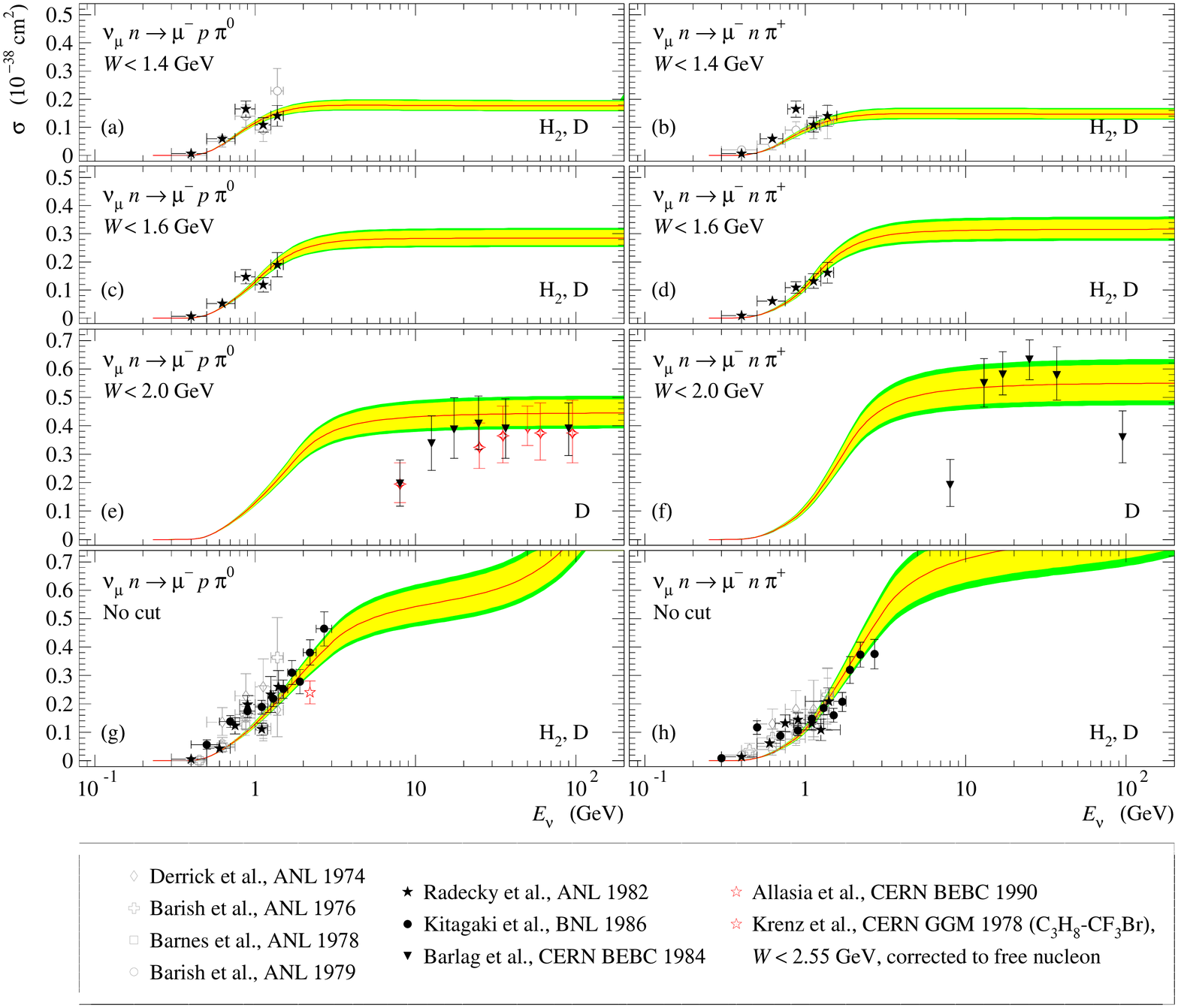}
  \caption{(Color online)
           Total cross sections
           of the reaction $\CCmnB$ (a), (c), (e), (g),
           $\CCmnC$ (b), (d), (f), (h)
           as functions of neutrino energy 
           predicted with $M_A$ and $f_\NRB$
           obtained from the global fits for H$_2$ and D data
           in comparison with the experimental data measured at
           ANL 1974 \cite{Derrick:1974ema},
               1976 \cite{Barish:1975bw},
               1978 \cite{Barnes:1978cs},
               1982 \cite{Radecky:1981fn},
               1979 \cite{Barish:1978pj}
           BNL 1986 \cite{Kitagaki:1986ct}, and
           CERN GGM 1978 \cite{Krenz:1977sw},
                BEBC 1984 \cite{Barlag:1984uga},
                     1990 \cite{Allasia:1990uy,Hawker:02}.
           Original experimental data measured at
           ANL 1982 \cite{Radecky:1981fn} for $W < 1.4$ GeV and with no cut for $W$;
           BNL 1986 \cite{Kitagaki:1986ct} with no cut for $W$,
           recalculated by Rodrigues~\etal\, \cite{Rodrigues:2016xjj}.
           Titles of the reactions, targets, and experimental cuts on $W$
           are given in the legends.
           Shaded bands around the curves show
           the joint uncertainty of the fitted value of $M_A$ and $f_\NRB$
           correspond to $1\sigma$ and $2\sigma$ errors.
           }
  \label{Fig:sRESCC_D_withNRB_1_103.1.31.301.h4_103.1.31.101.k2_2_BSc}
  \end{figure*}

  \begin{figure*}[htb!]
  \includegraphics[width=0.97\linewidth]{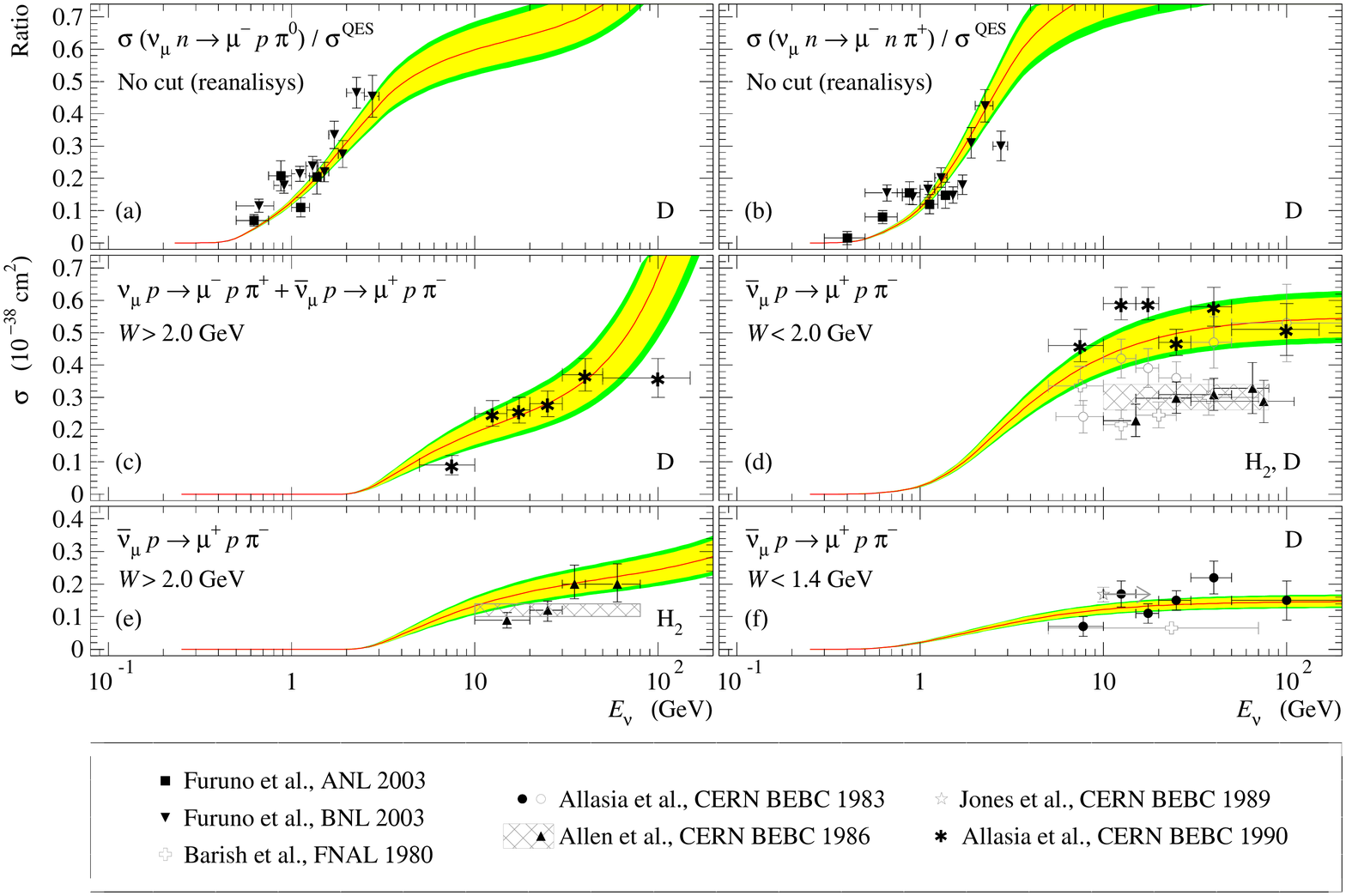}
  \caption{(Color online)
           Ratios of the total cross sections of the reactions
           $\CCmnB$ (a) and $\CCmnC$ (b) to the quasielastic reaction $\nu_{\mu}n\to\mu^-p$,
           sum of the reactions $\CCmnA$ and $\CCmaC$ (c), and
           total cross sections of the reaction $\CCmaC$ (d) -- (f)
           predicted with $M_A$ and $f_\NRB$
           obtained from the global fits for H$_2$ and D data
           in comparison with the experimental data
           measured at ANL and BNL \cite{Furuno:2003ng-proc,Sakuda:2002-KK,Sakuda:2003-KK},
           FNAL 1980 \cite{Barish:1979ny} (measured with cuts of $W < 1.9$ GeV and $W < 1.32$ GeV),
           CERN BEBC 1990 \cite{Allasia:1990uy},
                     1983 \cite{Allasia:1983qh},
                     1986 \cite{Allen:1985ti}, and 
                     1989 \cite{Jones:1989vt}.
           Titles of the reactions, targets, and experimental cuts on $W$
           are given in the legends.
           Notation for the solid curves and shaded bands is the same as in
           Fig. \ref{Fig:sRESCC_D_withNRB_1_103.1.31.301.h4_103.1.31.101.k2_2_BSc}.
           }
  \label{Fig:sRESCC_D_withNRB_2_103.1.31.301.h4_103.1.31.101.k2_2_BSc_101.3.31.301.2k_2_BBBA25}
  \end{figure*}

  \begin{figure*}[htb!]
  \includegraphics[width=0.97\linewidth]{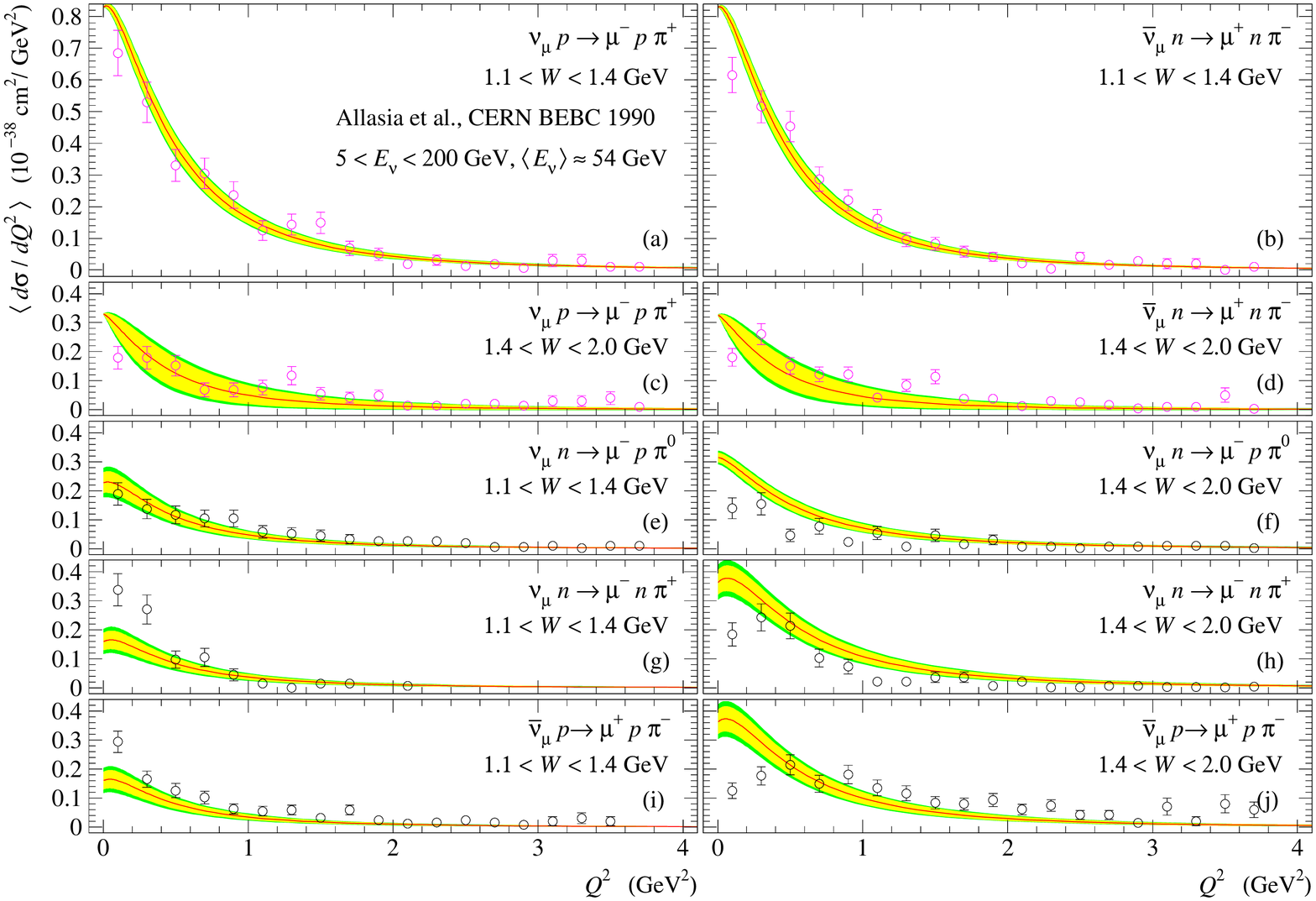}
  \caption{(Color online)
           Flux-weighted differential $Q^2$-dependent
           cross sections of the reactions
           $\CCmnA$ (a), (c), 
           $\CCmnB$ (e), (f),
           $\CCmnC$ (g), (h),
           $\CCmaA$ (b), (d), and
           $\CCmaC$ (i), (j)
           predicted with $M_A$ and $f_\NRB$
           obtained from the global fits for H$_2$ and D data
           in comparison with the experimental data 
           measured at CERN BEBC 1990 \cite{Allasia:1990uy}.
           Neutrino and antineutrino fluxes are borrowed from \cite{Allasia:1983dq}.
           Titles of the reactions, targets, and experimental cuts on $W$
           and neutrino energy are given in the legends.
           Inner and outer shaded bands around the curves
           in panels (a), (c), (b), (d) demonstrate
           the uncertainty of the fitted value of $M_A$
           corresponding to $1\sigma$ and $2\sigma$ errors, respectively.
           In the other panels the shaded bands demonstrate
           the joint uncertainty of the fitted value of $M_A$ and $f_\NRB$.
          }  
  \label{Fig:dsRESCC_dQ2_Allasia_BEBC90_103.1.31.301.h4_103.1.31.101.k2_2_BSc}
  \end{figure*}

  The most statistically reliable measurements
  of the total and differential cross sections,
  which were not superseded or reconsidered
  (due to increased statistics, revised of neutrino fluxes, etc.)
  in the subsequent reports of the same experimental groups,
  have been selected for our analysis.
  Further information and reanalysis of the original data
  are taken from the papers \cite{Furuno:2003ng-proc,Sakuda:2002-KK,Sakuda:2003-KK,Rodrigues:2016xjj,Hawker:02}.

  In the statistical analysis we use the experimental data
  on measuring the ratio of the total cross section
  of the resonance 1$\pi^+$-production reactions
  to the total cross section of the $\nu_{\mu}n\to\mu^-p$ reaction.
  The CC QES interactions on a bound nucleon in any kind of nuclei are calculated
  according to the phenomenological prescription
  of the effective running quasielastic axial-vector mass, $M_A^\text{run}$
  (see \cite{Kakorin:2020atz} and references therein).
  The parameters of $M_A^\text{run}$
  are obtained from the global fit to all available data
  on the $\nu_\mu$ and $\overline{\nu}_\mu$ cross sections of CC QES reactions.
  For neutrino reactions with deuterium and hydrogen,
  the running axial-vector mass is reduced to the current axial-vector mass of
  $M^\text{QES}_A = 1.008 \pm 0.025 (0.029)$~GeV.

  The process $\CCmnD$ was studied in experiments with deuterium targets at
  ANL 1979 \cite{Barish:1978pj} (nine data points),
  BNL 1986 \cite{Kitagaki:1986ct} (8), and
  CERN 1980 \cite{Allen:1980ti} (7)
  with the cuts of $W < 1.4$~GeV.
  The data are presented as 15 and 9 data points
  for the total and flux-averaged $Q^2$-dependent differential cross sections, respectively.
  The total and differential cross sections of the $\CCmnD$ reaction
  were obtained in ANL 1979 \cite{Barish:1978pj} experiment.
  We use the differential cross sections for the global fit
  and the total cross sections for the comparison only.
  The experimental data of the total cross sections
  of the reaction $\Delta^{++}$ resonance production were measured at
  ANL 1973 \cite{Campbell:1973wg} and
  FNAL 1978 \cite{Bell:1978rb,Bell:1978qu}
  and were obtained with no experimental cuts for $W$.
  The cross sections measured at FNAL 1978 \cite{Bell:1978rb,Bell:1978qu}
  are corrected by a factor of 1.1 due to the following reason.
  The particle fluxes from the broad-band horn beams at FNAL were not measured directly.
  Therefore, the results were normalized by obtaining the ratio $\CCmnD$ to all CC events and using
  the total cross section from FNAL 1977 \cite{Barish:1977ny} and CERN BEBC 1977 \cite{Bosetti:1977nd}
  parametrized as $\sigma\left({\nu}N\right)=\left(0.77-0.085\lg E_\nu\right)E_\nu\times 10^{-38}\,\text{cm}^2$.
  This parametrization underestimates the cross section to about 11\%
  (the estimate is made with the GRV98 \cite{Gluck:1998xa} parton distribution functions model).
  Actually, the papers \cite{Campbell:1973wg,Bell:1978rb,Bell:1978qu,Lerche:1978cp}
  do not contain clear information about of the limits for $W$.
  It is not obvious that the old experiments with small statistics of the neutrino-induced events
  could reliably identify the $\Delta^{++}$ resonance and background from other resonances.
  If the data of ANL 1973 \cite{Campbell:1973wg} and
  FNAL 1978 \cite{Bell:1978rb,Bell:1978qu}
  experiments were obtained for $W < 1.4$ GeV,
  the results of their measurement would be superseded in the subsequent papers.
  To avoid misleading interpretation of the data,
  we do not use these datasets for the global fits.

  The experimental data for the cross sections
  of the $\CCmnA$ reaction are obtained at
  ANL 1982 \cite{Radecky:1981fn} (23 data points),
  BNL 1986 \cite{Kitagaki:1986ct} (7),
  FNAL 1978 \cite{Bell:1978rb,Bell:1978qu} (2), and
  CERN BEBC 1984 \cite{Barlag:1984uga} (6),
            1986 \cite{Allen:1985ti} (9), and
            1990 \cite{Allasia:1990uy} (6)
  on deuterium targets for different values of the experimental cuts for $W$.
  The experiments to study the 1$\pi$-production reactions
  at ANL accelerator center were terminated in 1982 and not resumed but
  the data accumulated by ANL 1982 experiment \cite{Radecky:1981fn}
  are still the most important.
  The total cross sections were measured with 
  cuts for $W$ of 1.4, 1.6 GeV and with no cut for $W$.
  The total cross sections measured for $W < 1.4$ GeV and with no cut for $W$
  were revised in the paper \cite{Rodrigues:2016xjj}.
  For the global fit we use the revised total cross sections with no cut for $W$.
  Instead of the total cross sections measured for $W < 1.4$ GeV
  we include into the global fit the original flux-averaged $Q^2$-dependent cross sections.
  For the global fits, we use 16 data points
  recalculated from the number of events measured at
  ANL 1982 \cite{Radecky:1981fn} (7 data points) and
  BNL 1986 \cite{Kitagaki:1986ct} (9)
  to the total cross sections ratio of the $\CCmnA$
  and CC QES reactions \cite{Furuno:2003ng-proc,Sakuda:2002-KK,Sakuda:2003-KK}.
  In the experiment with the deuterium target CERN BEBC 1990 \cite{Allasia:1990uy},
  the flux-averaged $Q^2$-dependent differential cross sections
  were studied for special cuts for $W$.
  Unfortunately, the differential cross sections 
  cannot be used in the global fits
  because the original $\nu_\mu$ flux is unknown.
  The data measured at
  ANL \cite{Perkins:1975bj,%
            Barish:1975bw,%
            Barish:1978pj,%
            Barnes:1978cs},
  FNAL \cite{Ross:1978yu,%
             Barish:1979ny}, and
  CERN \cite{Jones:1989vt}
  are obsolete.
  The dataset consists of 69 data points
  satisfying our selection criteria of the experimental data.

  The data on the measurement of the total cross sections
  of the $\CCmaA$ reaction were obtained at
  CERN BEBC 1983 \cite{Allasia:1983qh} (6 data points) and
            1990 \cite{Allasia:1990uy} (6).
  The paper with preliminary results of the experiment \cite{Allasia:1983qh}
  contains the values of the total cross sections for the $W$ cuts of 1.4 and 2 GeV,
  while the final paper provides the data only for $W < 2$ GeV.
  We use the final and preliminary data for $W < 2$ GeV
  but not reconsidered data for $W < 1.4$ GeV,
  12 data points in total.
  As well as in the case of the $\CCmnA$ reaction,
  we cannot use the data on the flux-averaged cross sections
  measured at CERN BEBC 1990 \cite{Allasia:1990uy}.
  Thus, 12 data points are included into the global fit.

  The cross sections of the $\CCmnB$ reaction were measured at
  ANL 1982 \cite{Radecky:1981fn} (16 data points),
  BNL 1986 \cite{Kitagaki:1986ct} (10), and
  CERN BEBC 1984 \cite{Barlag:1984uga} (6)
  with hydrogen and deuterium targets
  for the cuts of $W < 1.4$, 1.6, 2.0 GeV and without a cut.
  The total cross sections measured at ANL \cite{Radecky:1981fn} and
  BNL 1986\cite{Kitagaki:1986ct} for the cut of $W < 1.4$ GeV and without a cut
  were revised by Rodrigues~\etal\, \cite{Rodrigues:2016xjj}.
  The total cross sections measured at CERN BEBC 1990 \cite{Allasia:1990uy}
  for the cut of $W < 2$ GeV were revised by Hawker \cite{Hawker:02}.
  In the global fit we use the revised data.
  The raw data on measuring the $Q^2$-distributions obtained at
  ANL \cite{Radecky:1981fn} and
  BNL \cite{Kitagaki:1986ct} with no cut for $W$
  were recalculated by Furuno~\etal\, \cite{Furuno:2003ng-proc,Sakuda:2002-KK,Sakuda:2003-KK}
  to the ratio of the total cross sections of the $\CCmnB$ to CC QES reactions.
  We include these 13 data points into the global fit.
  We cannot use the data on the flux-averaged cross sections
  measured at CERN BEBC 1990 \cite{Allasia:1990uy}
  for the same reason as in the case of the $\CCmnA$ and $\CCmaA$ reactions.
  The data measured at
  ANL \cite{Derrick:1974ema,%
            Barish:1975bw,%
            Barnes:1978cs,%
            Barish:1978pj} and
  CERN GGM 1977 \cite{Krenz:1977sw}
  are updated in recent publications of the same author groups.
  Thus, there are 45 data points suitable for the global fit.

  The cross sections of the $\CCmnC$ reaction were measured at
  ANL 1982 \cite{Radecky:1981fn} (16 data points),
  BNL 1986 \cite{Kitagaki:1986ct} (11), and
  CERN BEBC 1984 \cite{Barlag:1984uga} (6).
  The same author groups revised the original data
  (except for the data measured at CERN BEBC \cite{Allasia:1990uy}
  for $W < 2$ GeV).
  Hence, there are 33 data points for the total cross sections
  and 14 data points for the total cross section ratios
  of the $\CCmnC$ to CC QES reactions \cite{Furuno:2003ng-proc,Sakuda:2002-KK,Sakuda:2003-KK}.

  The cross sections of the $\CCmaC$ reaction were measured at
  FNAL 1980 \cite{Barish:1979ny}, and
  CERN BEBC 1983 \cite{Allasia:1983qh} (6 data points),
            1986 \cite{Allen:1985ti} (4), and
            1990 \cite{Allasia:1990uy} (6).
  Therefore, 16 data points can be used for global fits.

  We use six data points
  for the total cross section sums
  of the $\CCmnA$ and $\CCmaC$ reactions measured at
  CERN BEBC 1990 \cite{Allasia:1990uy}.
  The cross sections of the $\CCmnB$ and $\CCmnC$ reactions
  with a neutral particle in the final state cannot be analyzed
  in the same experiment with kinematic fitting and
  were recognized by topology of secondary particles.
  The problem of the event separation is more difficult than any others
  because the selection is estimated for proton or pion tracks at 
  the technical limit of effective bubble density.
  This special feature leads to ambiguity in the interpretation of a specific reaction.
  Due to this reason six data points are not included into the global fit.

  The data of flux-averaged $Q^2$- and $W$-distributions measured at 
  ANL 1980 \cite{Derrick:1980nr},
      1982 \cite{Radecky:1981fn},
  BNL 1980 \cite{Fanourakis:1980si},
      1986 \cite{Kitagaki:1986ct},
  FNAL 1980 \cite{Barish:1979ny},
       BEBC 1986 \cite{Allen:1985ti},
            1990 \cite{Allasia:1990uy},
  are used only for comparison and not for global fits.
  Absolute values and shape of the distributions
  as well as the flux-averaged differential cross sections depending on $Q^2$
  are primarily determined by the shape of neutrino fluxes
  used in data processing by the authors of the experiments.
  The differential distribution estimated at mean energy of the neutrino flux
  or calculated with a nonoriginal flux
  provides only a qualitative comparison with the measured distribution,
  i.e., approximate calculations of differential distributions
  are a source of uncontrolled systematic uncertainty
  for the results of the global fits.

  We avoid using the differential cross sections measured at
  CERN BEBC 1990 \cite{Allasia:1990uy}
  because the data points seem self-contradictory.
  The shapes of the cross sections for different reactions
  are inconsistent with the predictions by the RS approach.
  These data lead to an increase in $\chi^2$ values
  but do not to clarify the fitted parameters. 
  The data of this experiment are shown for a qualitative comparison only.

  There are no available experimental data for the total or
  differential cross sections of the $\CCmaB$ reaction
  measured on hydrogen or deuterium targets.


  \section{Results of the global fits}

  In our previous study \cite{Kuzmin:2006dh},
  the value of $M_A$ was obtained from the global likelihood analysis
  for total cross sections measured at
  ANL 1982 \cite{Radecky:1981fn}, 
  BNL 1986 \cite{Kitagaki:1986ct}, 
  FNAL 1978 \cite{Bell:1978qu},
       1980 \cite{Barish:1979ny}, 
  CERN GGM 1978 \cite{Lerche:1978cp,Krenz:1977sw},
           1979 \cite{Bolognese:1979gf},
  CERN BEBC 1980 \cite{Allen:1980ti},
            1983 \cite{Allasia:1983qh},
            1986 \cite{Allen:1985ti},
            1990 \cite{Allasia:1990uy}, and 
  IHEP SKAT \cite{Grabosch:1988gw}.
  The dataset included cross sections of the reactions
  measured on various nuclear targets
  with and without NRB contribution according to the RS model.
  That dataset contained 190 data points
  with 127 data points of the $\nu_\mu$ cross sections
  (66.84\% of the total number) and
  63 data points of the $\overline{\nu}_\mu$ cross sections (33.16\%).
  In the previous analysis we did not run a likelihood tune of NRB and
  used the values of the Breit-Wigner normalizations
  calculated numerically according to the formal definition.
  Also, the normalizations of the experimental data are not discussed.

  The value of the resonance axial-vector mass
  \[
  M_A = 1.18_{-0.07\,(0.08)}^{+0.07\,(0.09)}\,\text{GeV}
  \]
  was obtained with $\chi^2/\ndf = 186.6/(105-11) \approx 1.99$
  from the global fit for the cross sections
  of the $\CCmnD$, $\CCmnA$, and $\CCmaA$ reactions
  which do not require NRB in the RS approach
  and are measured in experiments on hydrogen and deuterium targets only.
  The dataset consists of 105 experimental data points
  with 93 data points of the $\nu_\mu$ cross sections
  (88.6\% of the total number) and
  12 data points of the $\overline{\nu}_\mu$ cross sections (11.4\%).
 
  The value of the adjustable parameter
  \[
  f_\NRB = 1.12 \pm 0.10\,(0.13)
  \] 
  was obtained with $\chi^2/\ndf = 428.4/(114-8) \approx 4.04$
  from the fit for the cross sections
  of the $\CCmnB$, $\CCmnC$, and $\CCmaC$ reactions
  which require the consideration of NRB in the RS approach.
  The dataset consists of 114 experimental data points
  with 92 points of the $\nu_\mu$ cross sections
  (81\% of the total number),
  16 data points of the $\overline{\nu}_\mu$ cross sections (14\%), and
  6 data points for sum of the $\nu_\mu$ and $\overline{\nu}_\mu$ cross sections (5\%).

  To illustrate the effect of normalizations of the Breit-Wigner distributions
  on the discussed tunable parameters, 
  we perform global fits using a more traditional choice of normalizations
  according to the previous version of KLN-BS model:
  $M_A = 1.12 \pm 0.06\,(0.08)$ GeV with $\chi^2/\ndf = 201.2/(105-11) \approx 2.12$ and
  $f_\NRB = 1.09 \pm 0.11\,(0.13)$ with $\chi^2/\ndf = 413.1/(114-8) \approx 3.90$.
  The parameters extracted from the new global and test fits as well as $\chi^2$-values
  are slightly different, which means that normalizations of distributions
  do not significantly affect the global fit for the model parameters.
  However, unreliable/unphysical constants should be avoided.

  Tables \ref{Tab:MA_RES_individual_fits_D_H_2_no_NRB} and 
  \ref{Tab:FB_RES_individual_fits_D_H_2_with_NRB} collect
  the values of $M_A$ and the adjustable parameter $f_\NRB$
  with the normalization factors obtained from the individual fits
  for each experimental datasets involved into the global fit.
  The values of the parameters for individual fits agree
  with each other within the errors.
  Figures
  \ref{Fig:sRESCC_D_withoutNRB_103.1.31.301.h4_2_BSc},
  \ref{Fig:sRESCC+dsRESCC_dQ2_D_withoutNRB_103.1.31.301.h4_2_BSc_101.3.31.301.2k_2_BBBA25}
  \ref{Fig:sRESCC_D_withNRB_1_103.1.31.301.h4_103.1.31.101.k2_2_BSc},
  \ref{Fig:sRESCC_D_withNRB_2_103.1.31.301.h4_103.1.31.101.k2_2_BSc_101.3.31.301.2k_2_BBBA25}, and
  \ref{Fig:dsRESCC_dQ2_Allasia_BEBC90_103.1.31.301.h4_103.1.31.101.k2_2_BSc}
  show the experimental data in comparison with the predicted cross sections
  for default values for $M_A$ and $f_\NRB$ parameters.
  In all Figs.
  the data points marked with open symbols
  are not included into the statistical analysis
  as well as the averaged cross sections
  shown with hatched rectangles.
  The experimental data are not multiplied by normalization factors
  obtained from the global fits.
  We do not include into the global fit
  the experimental data on the flux-weighted total cross sections
  shown in Fig.~\ref{Fig:sRESCC_D_withoutNRB_103.1.31.301.h4_2_BSc}(b),
  \ref{Fig:sRESCC+dsRESCC_dQ2_D_withoutNRB_103.1.31.301.h4_2_BSc_101.3.31.301.2k_2_BBBA25}(a), and
  \ref{Fig:sRESCC_D_withNRB_2_103.1.31.301.h4_103.1.31.101.k2_2_BSc_101.3.31.301.2k_2_BBBA25}(e),
  \ref{Fig:sRESCC_D_withNRB_2_103.1.31.301.h4_103.1.31.101.k2_2_BSc_101.3.31.301.2k_2_BBBA25}(d)
  as shaded areas.

  The ANL and BNL data
  measured for the $\nu_\mu p \to \mu^- p \pi^+$ reaction without cut for $W$
  were corrected for nuclear effects and FSI \cite{Nakamura:2018ntd}.
  The global fit with the corrected data leads to the same $M_A$.
  The point is that  the cross section weakly depends on the axial mass
  in the low neutrino energy range and corrections
  don't significantly change the data

  \section{Conclusions}

  First, we recommend clarifying the RS model and its extensions
  by getting rid of the normalizations of the Breit-Wigner distributions
  to avoid unphysical quantities in calculations.

  Next, the global fit of the experimental data 
  on the total and differential CC cross sections
  for the $\nu_\mu$ and $\overline{\nu}_\mu$ reactions
  of the 1$\pi$-production through the nucleon and baryon resonances,
  measured in experiments on hydrogen and deuterium targets,
  allows one to determine the phenomenological parameters
  of the resonance current axial-vector mass
  and the adjustable constant for fine-tuned noninterfering nonresonance background
  for the phenomenological models based on the RS approach.
  The axial mass extracted from the fit for only $\nu_\mu$ data yields the same result.
  The low number of the $\overline{\nu}_\mu$ data
  does not allow one to obtain a reliable result from the global fits;
  however the result does not contradict the global analysis.
  The cross sections predicted within the clarified KLN-BS model
  are in good agreement with the experimental data
  in the whole range of neutrino energy.
  The updated value of the axial mass is consistent with the previous one
  and describes well the data obtained in experiments with nuclear targets \cite{Kuzmin:2006dh}.
  Modern experiments with deuterium and hydrogen targets
  operating with the $\nu_\mu$ and $\overline{\nu}_\mu$ beams
  are needed for a more precise determination
  of the parameters of the model.

  Finally, we emphasize that
  the obtained resonance axial mass and adjustable nonresonance background
  are appropriate only for the RS model and its extensions 
  in which the traditional normalization of the Breit-Wigner factors is abandoned.
  The alternative models for describing the resonance 1$\pi$-production reactions
  may require a different value of the axial mass
  and tune for the background induced by nonresonance reactions.

 \section*{Acknowledgements}

  We are very grateful to Vadim~A.~Naumov for many useful discussions and
  for drawing our attention to the problem
  with calculating of the resonance Breit-Wigner normalization factors
  in the RS approach.
  We are very grateful to Satoshi Nakamura alerted us to that
  ANL and BNL data need to be corrected for nuclear effects and FSI.
  We thank Galina G. Sandukovskaya for proofreading of the text.
  Research of Igor D.~Kakorin
  has been supported by the Russian Science Foundation Grant No.~18-12-00271.

%



\begin{thebibliography}{92}%
\makeatletter
\providecommand \@ifxundefined [1]{%
 \@ifx{#1\undefined}
}%
\providecommand \@ifnum [1]{%
 \ifnum #1\expandafter \@firstoftwo
 \else \expandafter \@secondoftwo
 \fi
}%
\providecommand \@ifx [1]{%
 \ifx #1\expandafter \@firstoftwo
 \else \expandafter \@secondoftwo
 \fi
}%
\providecommand \natexlab [1]{#1}%
\providecommand \enquote  [1]{``#1''}%
\providecommand \bibnamefont  [1]{#1}%
\providecommand \bibfnamefont [1]{#1}%
\providecommand \citenamefont [1]{#1}%
\providecommand \href@noop [0]{\@secondoftwo}%
\providecommand \href [0]{\begingroup \@sanitize@url \@href}%
\providecommand \@href[1]{\@@startlink{#1}\@@href}%
\providecommand \@@href[1]{\endgroup#1\@@endlink}%
\providecommand \@sanitize@url [0]{\catcode `\\12\catcode `\$12\catcode
  `\&12\catcode `\#12\catcode `\^12\catcode `\_12\catcode `\%12\relax}%
\providecommand \@@startlink[1]{}%
\providecommand \@@endlink[0]{}%
\providecommand \url  [0]{\begingroup\@sanitize@url \@url }%
\providecommand \@url [1]{\endgroup\@href {#1}{\urlprefix }}%
\providecommand \urlprefix  [0]{URL }%
\providecommand \Eprint [0]{\href }%
\providecommand \doibase [0]{https://doi.org/}%
\providecommand \selectlanguage [0]{\@gobble}%
\providecommand \bibinfo  [0]{\@secondoftwo}%
\providecommand \bibfield  [0]{\@secondoftwo}%
\providecommand \translation [1]{[#1]}%
\providecommand \BibitemOpen [0]{}%
\providecommand \bibitemStop [0]{}%
\providecommand \bibitemNoStop [0]{.\EOS\space}%
\providecommand \EOS [0]{\spacefactor3000\relax}%
\providecommand \BibitemShut  [1]{\csname bibitem#1\endcsname}%
\let\auto@bib@innerbib\@empty
\bibitem [{\citenamefont {Fernandez-Martinez}\ and\ \citenamefont
  {Meloni}(2011)}]{FernandezMartinez:2010dm}%
  \BibitemOpen
  \bibfield  {author} {\bibinfo {author} {\bibfnamefont {E.}~\bibnamefont
  {Fernandez-Martinez}}\ and\ \bibinfo {author} {\bibfnamefont
  {D.}~\bibnamefont {Meloni}},\ }\bibfield  {title} {\bibinfo {title}
  {{Importance of nuclear effects in the measurement of neutrino oscillation
  parameters}},\ }\href {https://doi.org/10.1016/j.physletb.2011.02.043}
  {\bibfield  {journal} {\bibinfo  {journal} {Phys.\ Lett.\ B}\ }\textbf
  {\bibinfo {volume} {697}},\ \bibinfo {pages} {477} (\bibinfo {year}
  {2011})},\ \Eprint {https://arxiv.org/abs/1010.2329} {arXiv:1010.2329
  [hep-ph]} \BibitemShut {NoStop}%
\bibitem [{\citenamefont {Meloni}\ and\ \citenamefont
  {Martini}(2012)}]{Meloni:2012fq}%
  \BibitemOpen
  \bibfield  {author} {\bibinfo {author} {\bibfnamefont {D.}~\bibnamefont
  {Meloni}}\ and\ \bibinfo {author} {\bibfnamefont {M.}~\bibnamefont
  {Martini}},\ }\bibfield  {title} {\bibinfo {title} {{Revisiting the T2K data
  using different models for the neutrino-nucleus cross sections}},\ }\href
  {https://doi.org/10.1016/j.physletb.2012.08.007} {\bibfield  {journal}
  {\bibinfo  {journal} {Phys.\ Lett.\ B}\ }\textbf {\bibinfo {volume} {716}},\
  \bibinfo {pages} {186} (\bibinfo {year} {2012})},\ \Eprint
  {https://arxiv.org/abs/1203.3335} {arXiv:1203.3335 [hep-ph]} \BibitemShut
  {NoStop}%
\bibitem [{\citenamefont {Benhar}\ and\ \citenamefont
  {Rocco}(2013)}]{Benhar:2013oba}%
  \BibitemOpen
  \bibfield  {author} {\bibinfo {author} {\bibfnamefont {O.}~\bibnamefont
  {Benhar}}\ and\ \bibinfo {author} {\bibfnamefont {N.}~\bibnamefont {Rocco}},\
  }\bibfield  {title} {\bibinfo {title} {{Nuclear effects in neutrino
  interactions and their impact on the determination of oscillation
  parameters}},\ }\href {https://doi.org/10.1155/2013/912702} {\bibfield
  {journal} {\bibinfo  {journal} {Adv.\ High Energy Phys.}\ }\textbf {\bibinfo
  {volume} {2013}},\ \bibinfo {pages} {912702} (\bibinfo {year} {2013})},\
  \Eprint {https://arxiv.org/abs/1310.3869} {arXiv:1310.3869 [nucl-th]}
  \BibitemShut {NoStop}%
\bibitem [{\citenamefont {Coloma}\ and\ \citenamefont
  {Huber}(2013)}]{Coloma:2013rqa}%
  \BibitemOpen
  \bibfield  {author} {\bibinfo {author} {\bibfnamefont {P.}~\bibnamefont
  {Coloma}}\ and\ \bibinfo {author} {\bibfnamefont {P.}~\bibnamefont {Huber}},\
  }\bibfield  {title} {\bibinfo {title} {{Impact of nuclear effects on the
  extraction of neutrino oscillation parameters}},\ }\href
  {https://doi.org/10.1103/PhysRevLett.111.221802} {\bibfield  {journal}
  {\bibinfo  {journal} {Phys.\ Rev.\ Lett.}\ }\textbf {\bibinfo {volume}
  {111}},\ \bibinfo {pages} {221802} (\bibinfo {year} {2013})},\ \Eprint
  {https://arxiv.org/abs/1307.1243} {arXiv:1307.1243 [hep-ph]} \BibitemShut
  {NoStop}%
\bibitem [{\citenamefont {Coloma}\ \emph {et~al.}(2014)\citenamefont {Coloma},
  \citenamefont {Huber}, \citenamefont {Jen},\ and\ \citenamefont
  {Mariani}}]{Coloma:2013tba}%
  \BibitemOpen
  \bibfield  {author} {\bibinfo {author} {\bibfnamefont {P.}~\bibnamefont
  {Coloma}}, \bibinfo {author} {\bibfnamefont {P.}~\bibnamefont {Huber}},
  \bibinfo {author} {\bibfnamefont {C.-M.}\ \bibnamefont {Jen}},\ and\ \bibinfo
  {author} {\bibfnamefont {C.}~\bibnamefont {Mariani}},\ }\bibfield  {title}
  {\bibinfo {title} {{Neutrino-nucleus interaction models and their impact on
  oscillation analyses}},\ }\href {https://doi.org/10.1103/PhysRevD.89.073015}
  {\bibfield  {journal} {\bibinfo  {journal} {Phys.\ Rev.\ D}\ }\textbf
  {\bibinfo {volume} {89}},\ \bibinfo {pages} {073015} (\bibinfo {year}
  {2014})},\ \Eprint {https://arxiv.org/abs/1311.4506} {arXiv:1311.4506
  [hep-ph]} \BibitemShut {NoStop}%
\bibitem [{\citenamefont {Jen}\ \emph {et~al.}(2014)\citenamefont {Jen} \emph
  {et~al.}}]{Jen:2014aja}%
  \BibitemOpen
  \bibfield  {author} {\bibinfo {author} {\bibfnamefont {C.-M.}\ \bibnamefont
  {Jen}} \emph {et~al.},\ }\bibfield  {title} {\bibinfo {title} {{Numerical
  implementation of lepton-nucleus interactions and its effect on neutrino
  oscillation analysis}},\ }\href {https://doi.org/10.1103/PhysRevD.90.093004}
  {\bibfield  {journal} {\bibinfo  {journal} {Phys.\ Rev.\ D}\ }\textbf
  {\bibinfo {volume} {90}},\ \bibinfo {pages} {093004} (\bibinfo {year}
  {2014})},\ \Eprint {https://arxiv.org/abs/1402.6651} {arXiv:1402.6651
  [hep-ex]} \BibitemShut {NoStop}%
\bibitem [{\citenamefont {Ericson}\ and\ \citenamefont
  {Martini}(2015)}]{Ericson:2015cva}%
  \BibitemOpen
  \bibfield  {author} {\bibinfo {author} {\bibfnamefont {M.}~\bibnamefont
  {Ericson}}\ and\ \bibinfo {author} {\bibfnamefont {M.}~\bibnamefont
  {Martini}},\ }\bibfield  {title} {\bibinfo {title} {{Neutrino versus
  antineutrino cross sections and $CP$ violation}},\ }\href
  {https://doi.org/10.1103/PhysRevC.91.035501} {\bibfield  {journal} {\bibinfo
  {journal} {Phys.\ Rev.\ C}\ }\textbf {\bibinfo {volume} {91}},\ \bibinfo
  {pages} {035501} (\bibinfo {year} {2015})},\ \Eprint
  {https://arxiv.org/abs/1501.02442} {arXiv:1501.02442 [nucl-th]} \BibitemShut
  {NoStop}%
\bibitem [{\citenamefont {Ankowski}\ \emph {et~al.}(2016)\citenamefont
  {Ankowski}, \citenamefont {Benhar}, \citenamefont {Mariani},\ and\
  \citenamefont {Vagnoni}}]{Ankowski:2016bji}%
  \BibitemOpen
  \bibfield  {author} {\bibinfo {author} {\bibfnamefont {A.~M.}\ \bibnamefont
  {Ankowski}}, \bibinfo {author} {\bibfnamefont {O.}~\bibnamefont {Benhar}},
  \bibinfo {author} {\bibfnamefont {C.}~\bibnamefont {Mariani}},\ and\ \bibinfo
  {author} {\bibfnamefont {E.}~\bibnamefont {Vagnoni}},\ }\bibfield  {title}
  {\bibinfo {title} {{Effect of the $2p2h$ cross-section uncertainties on an
  analysis of neutrino oscillations}},\ }\href
  {https://doi.org/10.1103/PhysRevD.93.113004} {\bibfield  {journal} {\bibinfo
  {journal} {Phys.\ Rev.\ D}\ }\textbf {\bibinfo {volume} {93}},\ \bibinfo
  {pages} {113004} (\bibinfo {year} {2016})},\ \Eprint
  {https://arxiv.org/abs/1603.01072} {arXiv:1603.01072 [hep-ph]} \BibitemShut
  {NoStop}%
\bibitem [{\citenamefont {Ankowski}\ and\ \citenamefont
  {Mariani}(2017)}]{Ankowski:2016jdd}%
  \BibitemOpen
  \bibfield  {author} {\bibinfo {author} {\bibfnamefont {A.~M.}\ \bibnamefont
  {Ankowski}}\ and\ \bibinfo {author} {\bibfnamefont {C.}~\bibnamefont
  {Mariani}},\ }\bibfield  {title} {\bibinfo {title} {{Systematic uncertainties
  in long-baseline neutrino-oscillation experiments}},\ }\href
  {https://doi.org/10.1088/1361-6471/aa61b2} {\bibfield  {journal} {\bibinfo
  {journal} {J.\ Phys.\ G}\ }\textbf {\bibinfo {volume} {44}},\ \bibinfo
  {pages} {054001} (\bibinfo {year} {2017})},\ \Eprint
  {https://arxiv.org/abs/1609.00258} {arXiv:1609.00258 [hep-ph]} \BibitemShut
  {NoStop}%
\bibitem [{\citenamefont {Mosel}(2019)}]{Mosel:2019vhx}%
  \BibitemOpen
  \bibfield  {author} {\bibinfo {author} {\bibfnamefont {U.}~\bibnamefont
  {Mosel}},\ }\bibfield  {title} {\bibinfo {title} {{Neutrino event generators:
  foundation, status and future}},\ }\href
  {https://doi.org/10.1088/1361-6471/ab3830} {\bibfield  {journal} {\bibinfo
  {journal} {J.\ Phys.\ G}\ }\textbf {\bibinfo {volume} {46}},\ \bibinfo
  {pages} {113001} (\bibinfo {year} {2019})},\ \Eprint
  {https://arxiv.org/abs/1904.11506} {arXiv:1904.11506 [hep-ex]} \BibitemShut
  {NoStop}%
\bibitem [{\citenamefont {Kuzmin}\ \emph {et~al.}()\citenamefont {Kuzmin},
  \citenamefont {Lyubushkin},\ and\ \citenamefont {Naumov}}]{Kuzmin:2005bm}%
  \BibitemOpen
  \bibfield  {author} {\bibinfo {author} {\bibfnamefont {K.~S.}\ \bibnamefont
  {Kuzmin}}, \bibinfo {author} {\bibfnamefont {V.~V.}\ \bibnamefont
  {Lyubushkin}},\ and\ \bibinfo {author} {\bibfnamefont {V.~A.}\ \bibnamefont
  {Naumov}},\ }\href@noop {} {\bibinfo {title} {{How to sum contributions into
  the total charged-current neutrino-nucleon cross section}}},\ \Eprint
  {https://arxiv.org/abs/hep-ph/0511308} {hep-ph/0511308} \BibitemShut
  {NoStop}%
\bibitem [{\citenamefont {Kuzmin}\ \emph
  {et~al.}(2006{\natexlab{a}})\citenamefont {Kuzmin}, \citenamefont
  {Lyubushkin},\ and\ \citenamefont {Naumov}}]{Kuzmin:2006dt}%
  \BibitemOpen
  \bibfield  {author} {\bibinfo {author} {\bibfnamefont {K.~S.}\ \bibnamefont
  {Kuzmin}}, \bibinfo {author} {\bibfnamefont {V.~V.}\ \bibnamefont
  {Lyubushkin}},\ and\ \bibinfo {author} {\bibfnamefont {V.~A.}\ \bibnamefont
  {Naumov}},\ }\bibfield  {title} {\bibinfo {title} {{Fine-tuning parameters to
  describe the total charged-current neutrino-nucleon cross section}},\
  }\bibfield  {booktitle} {\emph {\bibinfo {booktitle} {{Proceedings of the 5th
  International Conference on Non-accelerator New Physics (NANP\,2005), Dubna,
  Russia, June 20--25, 2005}}},\ }\href
  {https://doi.org/10.1134/S1063778806110081} {\bibfield  {journal} {\bibinfo
  {journal} {Phys.\ Atom.\ Nucl.}\ }\textbf {\bibinfo {volume} {69}},\ \bibinfo
  {pages} {1857} (\bibinfo {year} {2006}{\natexlab{a}})}\BibitemShut {NoStop}%
\bibitem [{\citenamefont {Ravndal}(1973)}]{Ravndal:1973xx}%
  \BibitemOpen
  \bibfield  {author} {\bibinfo {author} {\bibfnamefont {F.}~\bibnamefont
  {Ravndal}},\ }\bibfield  {title} {\bibinfo {title} {{Weak production of
  nuclear resonances in a relativistic quark model}},\ }\href
  {https://doi.org/10.1007/BF02722789} {\bibfield  {journal} {\bibinfo
  {journal} {Nuovo Cimento A}\ }\textbf {\bibinfo {volume} {18}},\ \bibinfo
  {pages} {385} (\bibinfo {year} {1973})}\BibitemShut {NoStop}%
\bibitem [{\citenamefont {Rein}\ and\ \citenamefont
  {Sehgal}(1981)}]{Rein:1980wg}%
  \BibitemOpen
  \bibfield  {author} {\bibinfo {author} {\bibfnamefont {D.}~\bibnamefont
  {Rein}}\ and\ \bibinfo {author} {\bibfnamefont {L.~M.}\ \bibnamefont
  {Sehgal}},\ }\bibfield  {title} {\bibinfo {title} {{Neutrino-excitation of
  baryon resonances and single pion production}},\ }\href
  {https://doi.org/10.1016/0003-4916(81)90242-6} {\bibfield  {journal}
  {\bibinfo  {journal} {Ann.\ Phys.\ (N.\ Y.)}\ }\textbf {\bibinfo {volume}
  {133}},\ \bibinfo {pages} {79} (\bibinfo {year} {1981})}\BibitemShut
  {NoStop}%
\bibitem [{\citenamefont {Rein}(1987)}]{Rein:1987cb}%
  \BibitemOpen
  \bibfield  {author} {\bibinfo {author} {\bibfnamefont {D.}~\bibnamefont
  {Rein}},\ }\bibfield  {title} {\bibinfo {title} {{Angular distribution in
  neutrino induced single pion production processes}},\ }\href
  {https://doi.org/10.1007/BF01561054} {\bibfield  {journal} {\bibinfo
  {journal} {Z.\ Phys.\ C}\ }\textbf {\bibinfo {volume} {35}},\ \bibinfo
  {pages} {43} (\bibinfo {year} {1987})}\BibitemShut {NoStop}%
\bibitem [{\citenamefont {Feynman}\ \emph {et~al.}(1971)\citenamefont
  {Feynman}, \citenamefont {Kislinger},\ and\ \citenamefont
  {Ravndal}}]{Feynman:1971wr}%
  \BibitemOpen
  \bibfield  {author} {\bibinfo {author} {\bibfnamefont {R.~P.}\ \bibnamefont
  {Feynman}}, \bibinfo {author} {\bibfnamefont {M.}~\bibnamefont {Kislinger}},\
  and\ \bibinfo {author} {\bibfnamefont {F.}~\bibnamefont {Ravndal}},\
  }\bibfield  {title} {\bibinfo {title} {{Current matrix elements from a
  relativistic quark model}},\ }\href {https://doi.org/10.1103/PhysRevD.3.2706}
  {\bibfield  {journal} {\bibinfo  {journal} {Phys.\ Rev.\ D}\ }\textbf
  {\bibinfo {volume} {3}},\ \bibinfo {pages} {2706} (\bibinfo {year}
  {1971})}\BibitemShut {NoStop}%
\bibitem [{\citenamefont {Derrick}\ \emph {et~al.}(1978)\citenamefont {Derrick}
  \emph {et~al.}}]{Derrick:1978jz}%
  \BibitemOpen
  \bibfield  {author} {\bibinfo {author} {\bibfnamefont {M.}~\bibnamefont
  {Derrick}} \emph {et~al.},\ }\bibfield  {title} {\bibinfo {title} {{Simple
  charged-current channels in $\nu$-D$_2$ interactions}},\ }in\ \href@noop {}
  {\emph {\bibinfo {booktitle} {{Proceedings of the Topical Conference on
  Neutrino Physics at Accelerators, Oxford, England, UK, July 4--7, 1978}}}},\
  \bibinfo {editor} {edited by\ \bibinfo {editor} {\bibfnamefont {A.~G.}\
  \bibnamefont {Michette}}\ and\ \bibinfo {editor} {\bibfnamefont {P.~B.}\
  \bibnamefont {Renton}}}\ (\bibinfo  {publisher} {Science Research Council,
  Rutherford Laboratory},\ \bibinfo {address} {Chilton, England, UK},\ \bibinfo
  {year} {1978}),\ pp.\ \bibinfo {pages} {58--66}\BibitemShut {NoStop}%
\bibitem [{\citenamefont {Rollier}(1978)}]{Rollier:1978kr}%
  \BibitemOpen
  \bibfield  {author} {\bibinfo {author} {\bibfnamefont {M.}~\bibnamefont
  {Rollier}} (\bibinfo {collaboration} {Gargamelle Antineutrino
  Collaboration}),\ }\bibfield  {title} {\bibinfo {title} {{Recent results from
  the Gargamelle anti-neutrino propane experiment at the CERN PS}},\ }in\
  \href@noop {} {\emph {\bibinfo {booktitle} {{Proceedings of the Topical
  Conference on Neutrino Physics at Accelerators, Oxford, England, UK, July
  4--7, 1978}}}},\ \bibinfo {editor} {edited by\ \bibinfo {editor}
  {\bibfnamefont {A.~G.}\ \bibnamefont {Michette}}\ and\ \bibinfo {editor}
  {\bibfnamefont {P.~B.}\ \bibnamefont {Renton}}}\ (\bibinfo  {publisher}
  {Science Research Council, Rutherford Laboratory},\ \bibinfo {address}
  {Chilton, England, UK},\ \bibinfo {year} {1978}),\ pp.\ \bibinfo {pages}
  {68--74}\BibitemShut {NoStop}%
\bibitem [{\citenamefont {Dewit}(1978)}]{Dewit:1978}%
  \BibitemOpen
  \bibfield  {author} {\bibinfo {author} {\bibfnamefont {M.}~\bibnamefont
  {Dewit}},\ }\bibfield  {title} {\bibinfo {title} {{Experimental study of the
  reaction ${\nu}n\to\mu^-p$}},\ }in\ \href@noop {} {\emph {\bibinfo
  {booktitle} {{Proceedings of the Topical Conference on Neutrino Physics at
  Accelerators, Oxford, England, UK, July 4--7, 1978}}}},\ \bibinfo {editor}
  {edited by\ \bibinfo {editor} {\bibfnamefont {A.~G.}\ \bibnamefont
  {Michette}}\ and\ \bibinfo {editor} {\bibfnamefont {P.~B.}\ \bibnamefont
  {Renton}}}\ (\bibinfo  {publisher} {Science Research Council, Rutherford
  Laboratory},\ \bibinfo {address} {Chilton, England, UK},\ \bibinfo {year}
  {1978}),\ pp.\ \bibinfo {pages} {75--77}\BibitemShut {NoStop}%
\bibitem [{\citenamefont {Kuzmin}\ \emph
  {et~al.}(2006{\natexlab{b}})\citenamefont {Kuzmin}, \citenamefont
  {Lyubushkin},\ and\ \citenamefont {Naumov}}]{Kuzmin:2006dh}%
  \BibitemOpen
  \bibfield  {author} {\bibinfo {author} {\bibfnamefont {K.~S.}\ \bibnamefont
  {Kuzmin}}, \bibinfo {author} {\bibfnamefont {V.~V.}\ \bibnamefont
  {Lyubushkin}},\ and\ \bibinfo {author} {\bibfnamefont {V.~A.}\ \bibnamefont
  {Naumov}},\ }\bibfield  {title} {\bibinfo {title} {{Axial masses in
  quasielastic neutrino scattering and single-pion neutrinoproduction on
  nucleons and nuclei}},\ }\bibfield  {booktitle} {\emph {\bibinfo {booktitle}
  {{Proceedings of the 20th Max Born Symposium: Nuclear Effects in Neutrino
  Interactions, Wroclaw, Poland, December 7--10, 2005}}},\ }\href@noop {}
  {\bibfield  {journal} {\bibinfo  {journal} {Acta Phys.\ Polon.\ B}\ }\textbf
  {\bibinfo {volume} {37}},\ \bibinfo {pages} {2337} (\bibinfo {year}
  {2006}{\natexlab{b}})},\ \Eprint {https://arxiv.org/abs/hep-ph/0606184}
  {hep-ph/0606184} \BibitemShut {NoStop}%
\bibitem [{\citenamefont {Andreopoulos}\ \emph {et~al.}(2010)\citenamefont
  {Andreopoulos} \emph {et~al.}}]{Andreopoulos:2009rq}%
  \BibitemOpen
  \bibfield  {author} {\bibinfo {author} {\bibfnamefont {C.}~\bibnamefont
  {Andreopoulos}} \emph {et~al.} (\bibinfo {collaboration} {GENIE
  Collaboration}),\ }\bibfield  {title} {\bibinfo {title} {{The GENIE neutrino
  Monte Carlo generator}},\ }\href {https://doi.org/10.1016/j.nima.2009.12.009}
  {\bibfield  {journal} {\bibinfo  {journal} {Nucl.\ Instrum.\ Methods Phys.\
  Res.,\ Sect.\ A}\ }\textbf {\bibinfo {volume} {614}},\ \bibinfo {pages} {87}
  (\bibinfo {year} {2010})},\ \Eprint {https://arxiv.org/abs/0905.2517}
  {arXiv:0905.2517 [hep-ph]} \BibitemShut {NoStop}%
\bibitem [{\citenamefont {Andreopoulos}\ \emph {et~al.}()\citenamefont
  {Andreopoulos} \emph {et~al.}}]{Andreopoulos:2015wxa}%
  \BibitemOpen
  \bibfield  {author} {\bibinfo {author} {\bibfnamefont {C.}~\bibnamefont
  {Andreopoulos}} \emph {et~al.},\ }\bibfield  {title} {\bibinfo {title} {{The
  GENIE neutrino Monte Carlo generator: Physics and user manual, version
  2.10.0}},\ }\Eprint {https://arxiv.org/abs/1510.05494} {arXiv:1510.05494
  [hep-ph]} \BibitemShut {NoStop}%
\bibitem [{\citenamefont {McGivern}\ \emph {et~al.}(2016)\citenamefont
  {McGivern} \emph {et~al.}}]{McGivern:2016bwh}%
  \BibitemOpen
  \bibfield  {author} {\bibinfo {author} {\bibfnamefont {C.~L.}\ \bibnamefont
  {McGivern}} \emph {et~al.} (\bibinfo {collaboration} {MINER$\nu$A
  Collaboration}),\ }\bibfield  {title} {\bibinfo {title} {{Cross sections for
  $\nu_\mu$ and $\overline{\nu}_\mu$ induced pion production on hydrocarbon in
  the few GeV region using MINER$\nu$A}},\ }\href
  {https://doi.org/10.1103/PhysRevD.94.052005} {\bibfield  {journal} {\bibinfo
  {journal} {Phys.\ Rev.\ D}\ }\textbf {\bibinfo {volume} {94}},\ \bibinfo
  {pages} {052005} (\bibinfo {year} {2016})},\ \Eprint
  {https://arxiv.org/abs/1606.07127} {arXiv:1606.07127 [hep-ex]} \BibitemShut
  {NoStop}%
\bibitem [{\citenamefont {Adamson}\ \emph {et~al.}(2015)\citenamefont {Adamson}
  \emph {et~al.}}]{Adamson:2014pgc}%
  \BibitemOpen
  \bibfield  {author} {\bibinfo {author} {\bibfnamefont {P.}~\bibnamefont
  {Adamson}} \emph {et~al.} (\bibinfo {collaboration} {MINOS Collaboration}),\
  }\bibfield  {title} {\bibinfo {title} {{Study of quasielastic scattering
  using charged-current $\nu_\mu$-iron interactions in the MINOS near
  detector}},\ }\href {https://doi.org/10.1103/PhysRevD.91.012005} {\bibfield
  {journal} {\bibinfo  {journal} {Phys.\ Rev.\ D}\ }\textbf {\bibinfo {volume}
  {91}},\ \bibinfo {pages} {012005} (\bibinfo {year} {2015})},\ \Eprint
  {https://arxiv.org/abs/1410.8613} {arXiv:1410.8613 [hep-ex]} \BibitemShut
  {NoStop}%
\bibitem [{\citenamefont {Abe}\ \emph {et~al.}(2011)\citenamefont {Abe} \emph
  {et~al.}}]{Abe:2011ks}%
  \BibitemOpen
  \bibfield  {author} {\bibinfo {author} {\bibfnamefont {K.}~\bibnamefont
  {Abe}} \emph {et~al.} (\bibinfo {collaboration} {T2K Collaboration}),\
  }\bibfield  {title} {\bibinfo {title} {{The T2K Experiment}},\ }\href
  {https://doi.org/10.1016/j.nima.2011.06.067} {\bibfield  {journal} {\bibinfo
  {journal} {Nucl.\ Instrum.\ Methods Phys.\ Res.,\ Sect.\ A}\ }\textbf
  {\bibinfo {volume} {659}},\ \bibinfo {pages} {106} (\bibinfo {year}
  {2011})},\ \Eprint {https://arxiv.org/abs/1106.1238} {arXiv:1106.1238
  [physics.ins-det]} \BibitemShut {NoStop}%
\bibitem [{\citenamefont {Kuzmin}\ \emph {et~al.}(2004)\citenamefont {Kuzmin},
  \citenamefont {Lyubushkin},\ and\ \citenamefont {Naumov}}]{Kuzmin:2003ji}%
  \BibitemOpen
  \bibfield  {author} {\bibinfo {author} {\bibfnamefont {K.~S.}\ \bibnamefont
  {Kuzmin}}, \bibinfo {author} {\bibfnamefont {V.~V.}\ \bibnamefont
  {Lyubushkin}},\ and\ \bibinfo {author} {\bibfnamefont {V.~A.}\ \bibnamefont
  {Naumov}},\ }\bibfield  {title} {\bibinfo {title} {{Lepton polarization in
  neutrino nucleon interactions}},\ }\href
  {https://doi.org/10.1142/S0217732304016172} {\bibfield  {journal} {\bibinfo
  {journal} {Mod.\ Phys.\ Lett.\ A}\ }\textbf {\bibinfo {volume} {19}},\
  \bibinfo {pages} {2815} (\bibinfo {year} {2004})},\ \Eprint
  {https://arxiv.org/abs/hep-ph/0312107} {hep-ph/0312107} \BibitemShut
  {NoStop}%
\bibitem [{\citenamefont {Kuzmin}\ \emph {et~al.}(2005)\citenamefont {Kuzmin},
  \citenamefont {Lyubushkin},\ and\ \citenamefont {Naumov}}]{Kuzmin:2004ya}%
  \BibitemOpen
  \bibfield  {author} {\bibinfo {author} {\bibfnamefont {K.~S.}\ \bibnamefont
  {Kuzmin}}, \bibinfo {author} {\bibfnamefont {V.~V.}\ \bibnamefont
  {Lyubushkin}},\ and\ \bibinfo {author} {\bibfnamefont {V.~A.}\ \bibnamefont
  {Naumov}},\ }\bibfield  {title} {\bibinfo {title} {{Extended Rein-Sehgal
  model for tau lepton production}},\ }\bibfield  {booktitle} {\emph {\bibinfo
  {booktitle} {{Proceedings of the 3rd International Workshop on
  Neutrino-Nucleus Interactions in the Few GeV Region (NuInt\,2004), Assergi,
  Italy, March 17--21, 2004}}},\ }\href
  {https://doi.org/10.1016/j.nuclphysbps.2004.11.213} {\bibfield  {journal}
  {\bibinfo  {journal} {Nucl.\ Phys.\ B (Proc.\ Suppl.)}\ }\textbf {\bibinfo
  {volume} {139}},\ \bibinfo {pages} {158} (\bibinfo {year} {2005})},\ \Eprint
  {https://arxiv.org/abs/hep-ph/0408106} {hep-ph/0408106} \BibitemShut
  {NoStop}%
\bibitem [{\citenamefont {Berger}\ and\ \citenamefont
  {Sehgal}(2007)}]{Berger:2007rq}%
  \BibitemOpen
  \bibfield  {author} {\bibinfo {author} {\bibfnamefont {C.}~\bibnamefont
  {Berger}}\ and\ \bibinfo {author} {\bibfnamefont {L.~M.}\ \bibnamefont
  {Sehgal}},\ }\bibfield  {title} {\bibinfo {title} {{Lepton mass effects in
  single pion production by neutrinos}},\ }\href
  {https://doi.org/10.1103/PhysRevD.76.113004} {\bibfield  {journal} {\bibinfo
  {journal} {Phys.\ Rev.\ D}\ }\textbf {\bibinfo {volume} {76}},\ \bibinfo
  {pages} {113004} (\bibinfo {year} {2007})},\ \Eprint
  {https://arxiv.org/abs/0709.4378} {arXiv:0709.4378 [hep-ph]} \BibitemShut
  {NoStop}%
\bibitem [{\citenamefont {Graczyk}\ and\ \citenamefont
  {Sobczyk}(2008{\natexlab{a}})}]{Graczyk:2008zz}%
  \BibitemOpen
  \bibfield  {author} {\bibinfo {author} {\bibfnamefont {K.~M.}\ \bibnamefont
  {Graczyk}}\ and\ \bibinfo {author} {\bibfnamefont {J.~T.}\ \bibnamefont
  {Sobczyk}},\ }\bibfield  {title} {\bibinfo {title} {{Lepton mass effects in
  weak charged current single pion production}},\ }\href
  {https://doi.org/10.1103/PhysRevD.77.053003} {\bibfield  {journal} {\bibinfo
  {journal} {Phys.\ Rev.\ D}\ }\textbf {\bibinfo {volume} {77}},\ \bibinfo
  {pages} {053003} (\bibinfo {year} {2008}{\natexlab{a}})},\ \Eprint
  {https://arxiv.org/abs/0709.4634} {arXiv:0709.4634 [hep-ph]} \BibitemShut
  {NoStop}%
\bibitem [{\citenamefont {Graczyk}\ and\ \citenamefont
  {Sobczyk}(2008{\natexlab{b}})}]{Graczyk:2007bc}%
  \BibitemOpen
  \bibfield  {author} {\bibinfo {author} {\bibfnamefont {K.~M.}\ \bibnamefont
  {Graczyk}}\ and\ \bibinfo {author} {\bibfnamefont {J.~T.}\ \bibnamefont
  {Sobczyk}},\ }\bibfield  {title} {\bibinfo {title} {{Form factors in the
  quark resonance model}},\ }\href {https://doi.org/10.1103/PhysRevD.77.053001,
  10.1103/PhysRevD.79.079903} {\bibfield  {journal} {\bibinfo  {journal}
  {Phys.\ Rev.\ D}\ }\textbf {\bibinfo {volume} {77}},\ \bibinfo {pages}
  {053001} (\bibinfo {year} {2008}{\natexlab{b}})},\ \bibinfo {note} {[Erratum:
  {\it ibid}. {\bf 79}, 079903 (2008)]},\ \Eprint
  {https://arxiv.org/abs/0707.3561} {arXiv:0707.3561 [hep-ph]} \BibitemShut
  {NoStop}%
\bibitem [{\citenamefont
  {Kabirnezhad}(2017{\natexlab{a}})}]{Kabirnezhad:2017xzx}%
  \BibitemOpen
  \bibfield  {author} {\bibinfo {author} {\bibfnamefont {M.}~\bibnamefont
  {Kabirnezhad}},\ }\emph {\bibinfo {title} {{Improvement of single pion
  production for T2K experiment simulation tools}}},\ \href@noop {} {Ph.D.
  thesis},\ \bibinfo  {school} {National Center for Nuclear Research, Warsaw}
  (\bibinfo {year} {2017}{\natexlab{a}})\BibitemShut {NoStop}%
\bibitem [{\citenamefont {Hernandez}\ \emph {et~al.}(2007)\citenamefont
  {Hernandez}, \citenamefont {Nieves},\ and\ \citenamefont
  {Valverde}}]{Hernandez:2007qq}%
  \BibitemOpen
  \bibfield  {author} {\bibinfo {author} {\bibfnamefont {E.}~\bibnamefont
  {Hernandez}}, \bibinfo {author} {\bibfnamefont {J.}~\bibnamefont {Nieves}},\
  and\ \bibinfo {author} {\bibfnamefont {M.}~\bibnamefont {Valverde}},\
  }\bibfield  {title} {\bibinfo {title} {{Weak pion production off the
  nucleon}},\ }\href {https://doi.org/10.1103/PhysRevD.76.033005} {\bibfield
  {journal} {\bibinfo  {journal} {Phys.\ Rev.\ D}\ }\textbf {\bibinfo {volume}
  {76}},\ \bibinfo {pages} {033005} (\bibinfo {year} {2007})},\ \Eprint
  {https://arxiv.org/abs/hep-ph/0701149} {hep-ph/0701149} \BibitemShut
  {NoStop}%
\bibitem [{\citenamefont {Kabirnezhad}(2018)}]{Kabirnezhad:2017jmf}%
  \BibitemOpen
  \bibfield  {author} {\bibinfo {author} {\bibfnamefont {M.}~\bibnamefont
  {Kabirnezhad}},\ }\bibfield  {title} {\bibinfo {title} {{Single pion
  production in neutrino-nucleon interactions}},\ }\href
  {https://doi.org/10.1103/PhysRevD.97.013002} {\bibfield  {journal} {\bibinfo
  {journal} {Phys.\ Rev.\ D}\ }\textbf {\bibinfo {volume} {97}},\ \bibinfo
  {pages} {013002} (\bibinfo {year} {2018})},\ \Eprint
  {https://arxiv.org/abs/1711.02403} {arXiv:1711.02403 [hep-ph]} \BibitemShut
  {NoStop}%
\bibitem [{\citenamefont {Kabirnezhad}(2016)}]{Kabirnezhad:2016nwu}%
  \BibitemOpen
  \bibfield  {author} {\bibinfo {author} {\bibfnamefont {M.}~\bibnamefont
  {Kabirnezhad}},\ }\bibfield  {title} {\bibinfo {title} {{Single pion
  production in neutrino-nucleon interaction}},\ }\bibfield  {booktitle} {\emph
  {\bibinfo {booktitle} {{Proceedings of the 10th International Workshop on
  Neutrino-Nucleus Interactions in the Few GeV Region (NuInt\,2015), Osaka,
  Japan, November 16--21, 2015}}},\ }\href
  {https://doi.org/10.7566/JPSCP.12.010043} {\bibfield  {journal} {\bibinfo
  {journal} {JPS Conf.\ Proc.}\ }\textbf {\bibinfo {volume} {12}},\ \bibinfo
  {pages} {010043} (\bibinfo {year} {2016})}\BibitemShut {NoStop}%
\bibitem [{\citenamefont
  {Kabirnezhad}(2017{\natexlab{b}})}]{Kabirnezhad:2017dui}%
  \BibitemOpen
  \bibfield  {author} {\bibinfo {author} {\bibfnamefont {M.}~\bibnamefont
  {Kabirnezhad}},\ }\bibfield  {title} {\bibinfo {title} {{Single pion
  production in neutrino reactions}},\ }\bibfield  {booktitle} {\emph {\bibinfo
  {booktitle} {{Proceedings of the 27th International Conference on Neutrino
  Physics and Astrophysics (Neutrino\,2016), London, England, UK, July 4--9,
  2016}}},\ }\href {https://doi.org/10.1088/1742-6596/888/1/012122} {\bibfield
  {journal} {\bibinfo  {journal} {J.\ Phys.\ Conf.\ Ser.}\ }\textbf {\bibinfo
  {volume} {888}},\ \bibinfo {pages} {012122} (\bibinfo {year}
  {2017}{\natexlab{b}})}\BibitemShut {NoStop}%
\bibitem [{\citenamefont {Zyla}\ \emph {et~al.}(2020)\citenamefont {Zyla} \emph
  {et~al.}}]{Zyla:2020zbs}%
  \BibitemOpen
  \bibfield  {author} {\bibinfo {author} {\bibfnamefont {P.~A.}\ \bibnamefont
  {Zyla}} \emph {et~al.} (\bibinfo {collaboration} {Particle Data Group}),\
  }\bibfield  {title} {\bibinfo {title} {{Review of particle physics}},\ }\href
  {https://doi.org/10.1093/ptep/ptaa104} {\bibfield  {journal} {\bibinfo
  {journal} {Prog.\ Theor.\ Exp.\ Phys.}\ }\textbf {\bibinfo {volume} {2020}},\
  \bibinfo {pages} {083C01} (\bibinfo {year} {2020})}\BibitemShut {NoStop}%
\bibitem [{\citenamefont {Patrignani}\ \emph {et~al.}(2016)\citenamefont
  {Patrignani} \emph {et~al.}}]{Patrignani:2016xqp}%
  \BibitemOpen
  \bibfield  {author} {\bibinfo {author} {\bibfnamefont {C.}~\bibnamefont
  {Patrignani}} \emph {et~al.} (\bibinfo {collaboration} {Particle Data
  Group}),\ }\bibfield  {title} {\bibinfo {title} {{Review of Particle
  Physics}},\ }\href {https://doi.org/10.1088/1674-1137/40/10/100001}
  {\bibfield  {journal} {\bibinfo  {journal} {Chin.\ Phys.\ C}\ }\textbf
  {\bibinfo {volume} {40}},\ \bibinfo {pages} {100001} (\bibinfo {year}
  {2016})}\BibitemShut {NoStop}%
\bibitem [{Note1()}]{Note1}%
  \BibitemOpen
  \bibinfo {note} {The normalizations of the resonance Breit-Wigner
  distributions according to the previous version of KLN-BS model are 0.957 for
  $P_{33}(1234)$, 0.784 for $P_{11}(1450)$, 1.055 for $S_{31}(1620)$, 0.935 for
  $P_{33}(1640)$, 0.751 for $D_{33}(1730)$, 1.229 for $P_{31}(1920)$, 0.635 for
  $F_{35}(1920)$, 0.710 for $F_{37}(1950)$, 1.285 for $P_{33}(1960)$, 1.008 for
  $D_{13}(1525)$, 1.067 for $S_{11}(1540)$, 1.051 for $S_{11}(1640)$, 1.165 for
  $D_{13}(1670)$, 1.024 for $D_{15}(1680)$, 0.912 for $F_{15}(1680)$, 1.349 for
  $P_{11}(1710)$, 1.301 for $P_{13}(1740)$, and 0.619 for
  $F_{17}(1970)$.}\BibitemShut {Stop}%
\bibitem [{\citenamefont {Ahrens}\ \emph {et~al.}(1986)\citenamefont {Ahrens}
  \emph {et~al.}}]{Ahrens:1986ke}%
  \BibitemOpen
  \bibfield  {author} {\bibinfo {author} {\bibfnamefont {L.~A.}\ \bibnamefont
  {Ahrens}} \emph {et~al.},\ }\bibfield  {title} {\bibinfo {title}
  {{Determination of the neutrino fluxes in the Brookhaven wide-band beams}},\
  }\href {https://doi.org/10.1103/PhysRevD.34.75} {\bibfield  {journal}
  {\bibinfo  {journal} {Phys.\ Rev.\ D}\ }\textbf {\bibinfo {volume} {34}},\
  \bibinfo {pages} {75} (\bibinfo {year} {1986})}\BibitemShut {NoStop}%
\bibitem [{\citenamefont {James}(1994)}]{James:1994vla}%
  \BibitemOpen
  \bibfield  {author} {\bibinfo {author} {\bibfnamefont {F.}~\bibnamefont
  {James}},\ }\href@noop {} {\emph {\bibinfo {title} {{MINUIT -- function
  minimization and error analysis: Reference manual version 94.1}}}} (\bibinfo
  {year} {1994}),\ \bibinfo {note} {{CERN-D-506}}\BibitemShut {NoStop}%
\bibitem [{\citenamefont {James}\ and\ \citenamefont
  {Roos}(1975)}]{James:1975343}%
  \BibitemOpen
  \bibfield  {author} {\bibinfo {author} {\bibfnamefont {F.}~\bibnamefont
  {James}}\ and\ \bibinfo {author} {\bibfnamefont {M.}~\bibnamefont {Roos}},\
  }\bibfield  {title} {\bibinfo {title} {{MINUIT -- a system for function
  minimization and analysis of the parameter errors and correlations}},\ }\href
  {https://doi.org/10.1016/0010-4655(75)90039-9} {\bibfield  {journal}
  {\bibinfo  {journal} {Comput.\ Phys.\ Commun.}\ }\textbf {\bibinfo {volume}
  {10}},\ \bibinfo {pages} {343} (\bibinfo {year} {1975})}\BibitemShut
  {NoStop}%
\bibitem [{\citenamefont {Baranov}\ \emph {et~al.}(2016)\citenamefont
  {Baranov}, \citenamefont {Balashov}, \citenamefont {Kutovskiy},\ and\
  \citenamefont {Semenov}}]{Baranov:2016gvt}%
  \BibitemOpen
  \bibfield  {author} {\bibinfo {author} {\bibfnamefont {A.~V.}\ \bibnamefont
  {Baranov}}, \bibinfo {author} {\bibfnamefont {N.~A.}\ \bibnamefont
  {Balashov}}, \bibinfo {author} {\bibfnamefont {N.~A.}\ \bibnamefont
  {Kutovskiy}},\ and\ \bibinfo {author} {\bibfnamefont {R.~N.}\ \bibnamefont
  {Semenov}},\ }\bibfield  {title} {\bibinfo {title} {{JINR cloud
  infrastructure evolution}},\ }\href
  {https://doi.org/10.1134/S1547477116050071} {\bibfield  {journal} {\bibinfo
  {journal} {Phys.\ Part.\ Nucl.\ Lett.}\ }\textbf {\bibinfo {volume} {13}},\
  \bibinfo {pages} {672} (\bibinfo {year} {2016})}\BibitemShut {NoStop}%
\bibitem [{\citenamefont {Balashov}\ \emph {et~al.}(2018)\citenamefont
  {Balashov} \emph {et~al.}}]{Balashov:2018}%
  \BibitemOpen
  \bibfield  {author} {\bibinfo {author} {\bibfnamefont {N.~A.}\ \bibnamefont
  {Balashov}} \emph {et~al.},\ }\bibfield  {title} {\bibinfo {title} {{JINR
  cloud service for scientific and engineering computations}},\ }\href
  {https://doi.org/10.25559/SITITO.14.201801.061-072} {\bibfield  {journal}
  {\bibinfo  {journal} {Mod.\ Inf.\ Tech. and IT-Ed.}\ }\textbf {\bibinfo
  {volume} {14}},\ \bibinfo {pages} {61} (\bibinfo {year} {2018})}\BibitemShut
  {NoStop}%
\bibitem [{\citenamefont {Campbell}\ \emph {et~al.}(1973)\citenamefont
  {Campbell} \emph {et~al.}}]{Campbell:1973wg}%
  \BibitemOpen
  \bibfield  {author} {\bibinfo {author} {\bibfnamefont {J.}~\bibnamefont
  {Campbell}} \emph {et~al.},\ }\bibfield  {title} {\bibinfo {title} {{Study of
  the reaction ${\nu}p\to{\mu^-}p\pi^+$}},\ }\href
  {https://doi.org/10.1103/PhysRevLett.30.335} {\bibfield  {journal} {\bibinfo
  {journal} {Phys.\ Rev.\ Lett.}\ }\textbf {\bibinfo {volume} {30}},\ \bibinfo
  {pages} {335} (\bibinfo {year} {1973})}\BibitemShut {NoStop}%
\bibitem [{\citenamefont {Schreiner}\ and\ \citenamefont
  {Von~Hippel}(1973)}]{Schreiner:1973ka}%
  \BibitemOpen
  \bibfield  {author} {\bibinfo {author} {\bibfnamefont {P.~A.}\ \bibnamefont
  {Schreiner}}\ and\ \bibinfo {author} {\bibfnamefont {F.}~\bibnamefont
  {Von~Hippel}},\ }\bibfield  {title} {\bibinfo {title}
  {{${\nu}p\to\mu^-\Delta^{++}$: comparison with theory}},\ }\href
  {https://doi.org/10.1103/PhysRevLett.30.339} {\bibfield  {journal} {\bibinfo
  {journal} {Phys.\ Rev.\ Lett.}\ }\textbf {\bibinfo {volume} {30}},\ \bibinfo
  {pages} {339} (\bibinfo {year} {1973})}\BibitemShut {NoStop}%
\bibitem [{\citenamefont {Derrick}(1975)}]{Derrick:1974ema}%
  \BibitemOpen
  \bibfield  {author} {\bibinfo {author} {\bibfnamefont {M.}~\bibnamefont
  {Derrick}},\ }\bibfield  {title} {\bibinfo {title} {{Charged current neutrino
  reactions in the resonance region}},\ }in\ \href@noop {} {\emph {\bibinfo
  {booktitle} {{Proceedings of the 17th International Conference on High-Energy
  Physics (ICHEP\,1974), London, England, UK, July 1--10, 1974}}}},\ \bibinfo
  {editor} {edited by\ \bibinfo {editor} {\bibfnamefont {J.~R.}\ \bibnamefont
  {Smith}}}\ (\bibinfo  {publisher} {Science Research Council, Rutherford
  Laboratory},\ \bibinfo {address} {Chilton, England, UK},\ \bibinfo {year}
  {1975}),\ pp.\ \bibinfo {pages} {II.166--170}\BibitemShut {NoStop}%
\bibitem [{\citenamefont {Perkins}(1975)}]{Perkins:1975bj}%
  \BibitemOpen
  \bibfield  {author} {\bibinfo {author} {\bibfnamefont {D.~H.}\ \bibnamefont
  {Perkins}},\ }\bibfield  {title} {\bibinfo {title} {{Review of neutrino
  experiments}},\ }in\ \href@noop {} {\emph {\bibinfo {booktitle} {{Proceedings
  of the International Symposium on Lepton and Photon Interactions at High
  Energies, Leland Stanford Junior University, Stanford, California, USA,
  August 21--27, 1975}}}},\ \bibinfo {editor} {edited by\ \bibinfo {editor}
  {\bibfnamefont {T.~W.}\ \bibnamefont {Kirk}}}\ (\bibinfo  {publisher} {SLAC
  National Accelerator Laboratory},\ \bibinfo {address} {Stanford, California,
  USA},\ \bibinfo {year} {1975}),\ pp.\ \bibinfo {pages} {571--603}\BibitemShut
  {NoStop}%
\bibitem [{\citenamefont {Barish}\ \emph {et~al.}(1976)\citenamefont {Barish}
  \emph {et~al.}}]{Barish:1975bw}%
  \BibitemOpen
  \bibfield  {author} {\bibinfo {author} {\bibfnamefont {S.~J.}\ \bibnamefont
  {Barish}} \emph {et~al.},\ }\bibfield  {title} {\bibinfo {title} {{Study of
  the isospin properties of single-pion production by neutrinos}},\ }\href
  {https://doi.org/10.1103/PhysRevLett.36.179} {\bibfield  {journal} {\bibinfo
  {journal} {Phys.\ Rev.\ Lett.}\ }\textbf {\bibinfo {volume} {36}},\ \bibinfo
  {pages} {179} (\bibinfo {year} {1976})}\BibitemShut {NoStop}%
\bibitem [{\citenamefont {Barish}\ \emph
  {et~al.}(1977{\natexlab{a}})\citenamefont {Barish} \emph
  {et~al.}}]{Barish:1977qk}%
  \BibitemOpen
  \bibfield  {author} {\bibinfo {author} {\bibfnamefont {S.~J.}\ \bibnamefont
  {Barish}} \emph {et~al.},\ }\bibfield  {title} {\bibinfo {title} {{Study of
  neutrino interactions in hydrogen and deuterium: Description of the
  experiment and study of the reaction $\nu+d\to\mu^-+p+p_s$}},\ }\href
  {https://doi.org/10.1103/PhysRevD.16.3103} {\bibfield  {journal} {\bibinfo
  {journal} {Phys.\ Rev.\ D}\ }\textbf {\bibinfo {volume} {16}},\ \bibinfo
  {pages} {3103} (\bibinfo {year} {1977}{\natexlab{a}})}\BibitemShut {NoStop}%
\bibitem [{\citenamefont {Barnes}\ \emph {et~al.}(1978)\citenamefont {Barnes}
  \emph {et~al.}}]{Barnes:1978cs}%
  \BibitemOpen
  \bibfield  {author} {\bibinfo {author} {\bibfnamefont {V.~E.}\ \bibnamefont
  {Barnes}} \emph {et~al.},\ }\bibfield  {title} {\bibinfo {title} {{Study of
  the isospin properties of single-pion production by neutrinos}},\ }in\
  \href@noop {} {\emph {\bibinfo {booktitle} {{Proceedings of the 6th
  International Conference on Neutrino Physics (Neutrino\,1978), Purdue
  University, West Lafayette, Indiana, USA, April 28 -- May 2, 1978}}}},\
  \bibinfo {editor} {edited by\ \bibinfo {editor} {\bibfnamefont {E.~C.}\
  \bibnamefont {Fowler}}}\ (\bibinfo  {publisher} {Purdue University Press},\
  \bibinfo {address} {West Lafayette, Indiana, USA},\ \bibinfo {year} {1978}),\
  pp.\ \bibinfo {pages} {C56--C63}\BibitemShut {NoStop}%
\bibitem [{\citenamefont {Barish}\ \emph {et~al.}(1979)\citenamefont {Barish}
  \emph {et~al.}}]{Barish:1978pj}%
  \BibitemOpen
  \bibfield  {author} {\bibinfo {author} {\bibfnamefont {S.~J.}\ \bibnamefont
  {Barish}} \emph {et~al.},\ }\bibfield  {title} {\bibinfo {title} {{Study of
  neutrino interactions in hydrogen and deuterium. II. Inelastic
  charged-current reactions}},\ }\href
  {https://doi.org/10.1103/PhysRevD.19.2521} {\bibfield  {journal} {\bibinfo
  {journal} {Phys.\ Rev.\ D}\ }\textbf {\bibinfo {volume} {19}},\ \bibinfo
  {pages} {2521} (\bibinfo {year} {1979})}\BibitemShut {NoStop}%
\bibitem [{\citenamefont {Derrick}\ \emph {et~al.}(1980)\citenamefont {Derrick}
  \emph {et~al.}}]{Derrick:1980nr}%
  \BibitemOpen
  \bibfield  {author} {\bibinfo {author} {\bibfnamefont {M.}~\bibnamefont
  {Derrick}} \emph {et~al.},\ }\bibfield  {title} {\bibinfo {title} {{Study of
  the reaction ${\nu}n\to{\nu}p\pi^-$}},\ }\href
  {https://doi.org/10.1016/0370-2693(80)90283-X, 10.1016/0370-2693(80)90191-4}
  {\bibfield  {journal} {\bibinfo  {journal} {Phys.\ Lett.}\ }\textbf {\bibinfo
  {volume} {92\,B}},\ \bibinfo {pages} {363} (\bibinfo {year} {1980})},\
  \bibinfo {note} {[Erratum: {\it ibid}.\ {\bf 95\,B}, 461 (1980)]}\BibitemShut
  {NoStop}%
\bibitem [{\citenamefont {Derrick}\ \emph {et~al.}(1981)\citenamefont {Derrick}
  \emph {et~al.}}]{Derrick:1980xw}%
  \BibitemOpen
  \bibfield  {author} {\bibinfo {author} {\bibfnamefont {M.}~\bibnamefont
  {Derrick}} \emph {et~al.},\ }\bibfield  {title} {\bibinfo {title} {{Study of
  single pion production by weak neutral currents in low-energy $\nu d$
  interactions}},\ }\href {https://doi.org/10.1103/PhysRevD.23.569} {\bibfield
  {journal} {\bibinfo  {journal} {Phys.\ Rev.\ D}\ }\textbf {\bibinfo {volume}
  {23}},\ \bibinfo {pages} {569} (\bibinfo {year} {1981})}\BibitemShut
  {NoStop}%
\bibitem [{\citenamefont {Radecky}\ \emph {et~al.}(1982)\citenamefont {Radecky}
  \emph {et~al.}}]{Radecky:1981fn}%
  \BibitemOpen
  \bibfield  {author} {\bibinfo {author} {\bibfnamefont {G.~M.}\ \bibnamefont
  {Radecky}} \emph {et~al.},\ }\bibfield  {title} {\bibinfo {title} {{Study of
  single-pion production by weak charged currents in low-energy ${\nu}d$
  interactions}},\ }\href {https://doi.org/10.1103/PhysRevD.25.1161,
  10.1103/PhysRevD.26.3297} {\bibfield  {journal} {\bibinfo  {journal} {Phys.\
  Rev.\ D}\ }\textbf {\bibinfo {volume} {25}},\ \bibinfo {pages} {1161}
  (\bibinfo {year} {1982})},\ \bibinfo {note} {[Erratum: {\it ibid}. {\bf 26},
  3297 (1982)]}\BibitemShut {NoStop}%
\bibitem [{\citenamefont {Lee}\ \emph {et~al.}(1977)\citenamefont {Lee} \emph
  {et~al.}}]{Lee:1976wr}%
  \BibitemOpen
  \bibfield  {author} {\bibinfo {author} {\bibfnamefont {W.-Y.}\ \bibnamefont
  {Lee}} \emph {et~al.},\ }\bibfield  {title} {\bibinfo {title} {{Single pion
  production in neutrino and anti-neutrino reactions}},\ }\href
  {https://doi.org/10.1103/PhysRevLett.38.202} {\bibfield  {journal} {\bibinfo
  {journal} {Phys.\ Rev.\ Lett.}\ }\textbf {\bibinfo {volume} {38}},\ \bibinfo
  {pages} {202} (\bibinfo {year} {1977})}\BibitemShut {NoStop}%
\bibitem [{\citenamefont {Fanourakis}\ \emph {et~al.}(1980)\citenamefont
  {Fanourakis} \emph {et~al.}}]{Fanourakis:1980si}%
  \BibitemOpen
  \bibfield  {author} {\bibinfo {author} {\bibfnamefont {G.~K.}\ \bibnamefont
  {Fanourakis}} \emph {et~al.},\ }\bibfield  {title} {\bibinfo {title} {{Study
  of low-energy anti-neutrino interactions on protons}},\ }\href
  {https://doi.org/10.1103/PhysRevD.21.562} {\bibfield  {journal} {\bibinfo
  {journal} {Phys.\ Rev.\ D}\ }\textbf {\bibinfo {volume} {21}},\ \bibinfo
  {pages} {562} (\bibinfo {year} {1980})}\BibitemShut {NoStop}%
\bibitem [{\citenamefont {Baker}\ \emph {et~al.}(1981)\citenamefont {Baker}
  \emph {et~al.}}]{Baker:1980pj}%
  \BibitemOpen
  \bibfield  {author} {\bibinfo {author} {\bibfnamefont {N.~J.}\ \bibnamefont
  {Baker}} \emph {et~al.},\ }\bibfield  {title} {\bibinfo {title} {{Study of
  the isospin structure of single pion production in charged current neutrino
  interactions}},\ }\href {https://doi.org/10.1103/PhysRevD.23.2495} {\bibfield
   {journal} {\bibinfo  {journal} {Phys.\ Rev.\ D}\ }\textbf {\bibinfo {volume}
  {23}},\ \bibinfo {pages} {2495} (\bibinfo {year} {1981})}\BibitemShut
  {NoStop}%
\bibitem [{\citenamefont {Baker}\ \emph {et~al.}(1983)\citenamefont {Baker}
  \emph {et~al.}}]{Baker:1983qc}%
  \BibitemOpen
  \bibfield  {author} {\bibinfo {author} {\bibfnamefont {N.~J.}\ \bibnamefont
  {Baker}} \emph {et~al.},\ }\bibfield  {title} {\bibinfo {title} {{Exclusive
  neutral-current reaction $\nu_{\mu}n\to\nu_{\mu}p\pi^-$ in the BNL 7-foot
  deuterium bubble chamber}},\ }\href
  {https://doi.org/10.1103/PhysRevD.28.2900} {\bibfield  {journal} {\bibinfo
  {journal} {Phys.\ Rev.\ D}\ }\textbf {\bibinfo {volume} {28}},\ \bibinfo
  {pages} {2900} (\bibinfo {year} {1983})}\BibitemShut {NoStop}%
\bibitem [{\citenamefont {Kitagaki}\ \emph {et~al.}(1986)\citenamefont
  {Kitagaki} \emph {et~al.}}]{Kitagaki:1986ct}%
  \BibitemOpen
  \bibfield  {author} {\bibinfo {author} {\bibfnamefont {T.}~\bibnamefont
  {Kitagaki}} \emph {et~al.},\ }\bibfield  {title} {\bibinfo {title} {{Charged
  current exclusive pion production in neutrino deuterium interactions}},\
  }\href {https://doi.org/10.1103/PhysRevD.34.2554} {\bibfield  {journal}
  {\bibinfo  {journal} {Phys.\ Rev.\ D}\ }\textbf {\bibinfo {volume} {34}},\
  \bibinfo {pages} {2554} (\bibinfo {year} {1986})}\BibitemShut {NoStop}%
\bibitem [{\citenamefont {Ross}(1978)}]{Ross:1978yu}%
  \BibitemOpen
  \bibfield  {author} {\bibinfo {author} {\bibfnamefont {R.~T.}\ \bibnamefont
  {Ross}} (\bibinfo {collaboration} {Fermilab--Hawaii--Berkeley--Michigan
  Collaboration}),\ }\bibfield  {title} {\bibinfo {title} {{A study of the
  reaction ${\nu}p\to\mu^-p\pi^+$}},\ }in\ \href@noop {} {\emph {\bibinfo
  {booktitle} {{Proceedings of the 6th International Conference on Neutrino
  Physics (Neutrino\,1978), Purdue University, West Lafayette, Indiana, USA,
  April 28 -- May 2, 1978}}}},\ \bibinfo {editor} {edited by\ \bibinfo {editor}
  {\bibfnamefont {E.~C.}\ \bibnamefont {Fowler}}}\ (\bibinfo  {publisher}
  {Purdue University Press},\ \bibinfo {address} {West Lafayette, Indiana,
  USA},\ \bibinfo {year} {1978}),\ pp.\ \bibinfo {pages} {929--938}\BibitemShut
  {NoStop}%
\bibitem [{\citenamefont {Bell}\ \emph
  {et~al.}(1978{\natexlab{a}})\citenamefont {Bell} \emph
  {et~al.}}]{Bell:1978rb}%
  \BibitemOpen
  \bibfield  {author} {\bibinfo {author} {\bibfnamefont {J.}~\bibnamefont
  {Bell}} \emph {et~al.},\ }\bibfield  {title} {\bibinfo {title} {{Study of the
  reaction ${\nu}p\to\mu^-\Delta^{++}$ at high energies and comparisons with
  theory}},\ }\href {https://doi.org/10.1103/PhysRevLett.41.1012} {\bibfield
  {journal} {\bibinfo  {journal} {Phys.\ Rev.\ Lett.}\ }\textbf {\bibinfo
  {volume} {41}},\ \bibinfo {pages} {1012} (\bibinfo {year}
  {1978}{\natexlab{a}})}\BibitemShut {NoStop}%
\bibitem [{\citenamefont {Bell}\ \emph
  {et~al.}(1978{\natexlab{b}})\citenamefont {Bell} \emph
  {et~al.}}]{Bell:1978qu}%
  \BibitemOpen
  \bibfield  {author} {\bibinfo {author} {\bibfnamefont {J.}~\bibnamefont
  {Bell}} \emph {et~al.},\ }\bibfield  {title} {\bibinfo {title}
  {{Cross-section measurements for the reactions ${\nu}p\to\mu^-p\pi^+$ and
  ${\nu}p\to\mu^-{p}K^+$ at high energies}},\ }\href
  {https://doi.org/10.1103/PhysRevLett.41.1008} {\bibfield  {journal} {\bibinfo
   {journal} {Phys.\ Rev.\ Lett.}\ }\textbf {\bibinfo {volume} {41}},\ \bibinfo
  {pages} {1008} (\bibinfo {year} {1978}{\natexlab{b}})}\BibitemShut {NoStop}%
\bibitem [{\citenamefont {Barish}\ \emph {et~al.}(1980)\citenamefont {Barish}
  \emph {et~al.}}]{Barish:1979ny}%
  \BibitemOpen
  \bibfield  {author} {\bibinfo {author} {\bibfnamefont {S.~J.}\ \bibnamefont
  {Barish}} \emph {et~al.},\ }\bibfield  {title} {\bibinfo {title} {{Study of
  the reaction $\overline{\nu}p\to\mu^+p\pi^+$}},\ }\href
  {https://doi.org/10.1016/0370-2693(80)90684-X} {\bibfield  {journal}
  {\bibinfo  {journal} {Phys.\ Lett.\ B}\ }\textbf {\bibinfo {volume} {91}},\
  \bibinfo {pages} {161} (\bibinfo {year} {1980})}\BibitemShut {NoStop}%
\bibitem [{\citenamefont {Franzinetti}(1966)}]{Franzinetti:276222}%
  \BibitemOpen
  \bibfield  {author} {\bibinfo {author} {\bibfnamefont {C.}~\bibnamefont
  {Franzinetti}},\ }\href {https://doi.org/10.5170/CERN-1966-013} {\emph
  {\bibinfo {title} {{Neutrino interactions in the CERN heavy-liquid
  bubble-chamber}}}},\ \bibinfo {type} {Tech. Rep.}\ \bibinfo {number} {CERN
  Yellow Report No.\,66-13}\ (\bibinfo {year} {1966})\BibitemShut {NoStop}%
\bibitem [{\citenamefont {Young}(1967)}]{Young:1967ud}%
  \BibitemOpen
  \bibfield  {author} {\bibinfo {author} {\bibfnamefont {E.~C.~M.}\
  \bibnamefont {Young}},\ }\href {https://doi.org/10.5170/CERN-1967-012} {\emph
  {\bibinfo {title} {{High-energy neutrino interactions}}}},\ \bibinfo {type}
  {Tech. Rep.}\ \bibinfo {number} {CERN Yellow Report No.\,67--12}\ (\bibinfo
  {year} {1967})\BibitemShut {NoStop}%
\bibitem [{\citenamefont {Budagov}\ \emph {et~al.}(1969)\citenamefont {Budagov}
  \emph {et~al.}}]{Budagov:1969pw}%
  \BibitemOpen
  \bibfield  {author} {\bibinfo {author} {\bibfnamefont {I.}~\bibnamefont
  {Budagov}} \emph {et~al.},\ }\bibfield  {title} {\bibinfo {title} {{Single
  pion production by neutrinos on free protons}},\ }\href
  {https://doi.org/10.1016/0370-2693(69)90041-0} {\bibfield  {journal}
  {\bibinfo  {journal} {Phys.\ Lett.}\ }\textbf {\bibinfo {volume} {29\,B}},\
  \bibinfo {pages} {524} (\bibinfo {year} {1969})}\BibitemShut {NoStop}%
\bibitem [{\citenamefont {Hasert}\ \emph {et~al.}(1975)\citenamefont {Hasert}
  \emph {et~al.}}]{Hasert:1975sv}%
  \BibitemOpen
  \bibfield  {author} {\bibinfo {author} {\bibfnamefont {F.~J.}\ \bibnamefont
  {Hasert}} \emph {et~al.},\ }\bibfield  {title} {\bibinfo {title} {{Neutral
  pion production by weak neutral currents in neutrino and antineutrino
  reactions}},\ }\href {https://doi.org/10.1016/0370-2693(75)90352-4}
  {\bibfield  {journal} {\bibinfo  {journal} {Phys.\ Lett.}\ }\textbf {\bibinfo
  {volume} {59\,B}},\ \bibinfo {pages} {485} (\bibinfo {year}
  {1975})}\BibitemShut {NoStop}%
\bibitem [{\citenamefont {Schmid}(1978)}]{Schmid:1978yt}%
  \BibitemOpen
  \bibfield  {author} {\bibinfo {author} {\bibfnamefont {P.}~\bibnamefont
  {Schmid}} (\bibinfo {collaboration} {Aachen--Bonn--CERN--Munich--Oxford
  Collaboration}),\ }\bibfield  {title} {\bibinfo {title} {{First results from
  a ${\nu}p$ experiment at CERN}},\ }\bibfield  {booktitle} {\emph {\bibinfo
  {booktitle} {{Proceedings of the 6th International Conference on Neutrino
  Physics (Neutrino\,1978), Purdue University, West Lafayette, Indiana, USA,
  April 28 -- May 2, 1978}}},\ }\href@noop {} {\bibfield  {journal} {\bibinfo
  {journal} {Conf.\ Proc.}\ }\textbf {\bibinfo {volume} {C780428}},\ \bibinfo
  {pages} {939} (\bibinfo {year} {1978})}\BibitemShut {NoStop}%
\bibitem [{\citenamefont {Krenz}\ \emph {et~al.}(1978)\citenamefont {Krenz}
  \emph {et~al.}}]{Krenz:1977sw}%
  \BibitemOpen
  \bibfield  {author} {\bibinfo {author} {\bibfnamefont {W.}~\bibnamefont
  {Krenz}} \emph {et~al.} (\bibinfo {collaboration} {Gargamelle Neutrino
  Propane, Aachen--Brussels--CERN--Ecole Poly--Orsay--Padua Collaborations}),\
  }\bibfield  {title} {\bibinfo {title} {{Experimental study of exclusive
  one-pion production in all neutrino-induced neutral current channels}},\
  }\href {https://doi.org/10.1016/0550-3213(78)90213-4} {\bibfield  {journal}
  {\bibinfo  {journal} {Nucl.\ Phys.\ B}\ }\textbf {\bibinfo {volume} {135}},\
  \bibinfo {pages} {45} (\bibinfo {year} {1978})}\BibitemShut {NoStop}%
\bibitem [{\citenamefont {Lerche}\ \emph {et~al.}(1978)\citenamefont {Lerche}
  \emph {et~al.}}]{Lerche:1978cp}%
  \BibitemOpen
  \bibfield  {author} {\bibinfo {author} {\bibfnamefont {W.}~\bibnamefont
  {Lerche}} \emph {et~al.},\ }\bibfield  {title} {\bibinfo {title}
  {{Experimental study of the reaction $\nu_{\mu}p\to\mu^-p\pi^+$ (Gargamelle
  Neutrino Propane experiment)}},\ }\href
  {https://doi.org/10.1016/0370-2693(78)90499-9} {\bibfield  {journal}
  {\bibinfo  {journal} {Phys.\ Lett.}\ }\textbf {\bibinfo {volume} {78\,B}},\
  \bibinfo {pages} {510} (\bibinfo {year} {1978})}\BibitemShut {NoStop}%
\bibitem [{\citenamefont {Pohl}(1978)}]{Pohl:1978pi}%
  \BibitemOpen
  \bibfield  {author} {\bibinfo {author} {\bibfnamefont {M.}~\bibnamefont
  {Pohl}} (\bibinfo {collaboration} {Aachen--Brussels--CERN--Ecole
  Polytechnique--Orsay--Padua Collaboration}),\ }\bibfield  {title} {\bibinfo
  {title} {{Experimental study of one pion production by the weak charged
  current}},\ }in\ \href@noop {} {\emph {\bibinfo {booktitle} {{Proceedings of
  the Topical Conference on Neutrino Physics at Accelerators, Oxford, England,
  UK, July 4--7, 1978}}}},\ \bibinfo {editor} {edited by\ \bibinfo {editor}
  {\bibfnamefont {A.~G.}\ \bibnamefont {Michette}}\ and\ \bibinfo {editor}
  {\bibfnamefont {P.~B.}\ \bibnamefont {Renton}}}\ (\bibinfo  {publisher}
  {Science Research Council, Rutherford Laboratory},\ \bibinfo {address}
  {Chilton, England, UK},\ \bibinfo {year} {1978}),\ pp.\ \bibinfo {pages}
  {78--82}\BibitemShut {NoStop}%
\bibitem [{\citenamefont {Pohl}\ \emph {et~al.}(1979)\citenamefont {Pohl} \emph
  {et~al.}}]{Pohl:1979fw}%
  \BibitemOpen
  \bibfield  {author} {\bibinfo {author} {\bibfnamefont {M.}~\bibnamefont
  {Pohl}} \emph {et~al.},\ }\bibfield  {title} {\bibinfo {title} {{Experimental
  study of single pion production in charged current neutrino interactions}},\
  }\href {https://doi.org/10.1007/BF02725470} {\bibfield  {journal} {\bibinfo
  {journal} {Lett.\ Nuovo Cimento}\ }\textbf {\bibinfo {volume} {24}},\
  \bibinfo {pages} {540} (\bibinfo {year} {1979})}\BibitemShut {NoStop}%
\bibitem [{\citenamefont {Bolognese}(1978)}]{Bolognese:1978yz}%
  \BibitemOpen
  \bibfield  {author} {\bibinfo {author} {\bibfnamefont {T.}~\bibnamefont
  {Bolognese}},\ }\emph {\bibinfo {title} {{Study of antineutrino interactions
  with charged current pion production}}},\ \href@noop {} {Ph.D. thesis},\
  \bibinfo  {school} {Strasbourg, CRN} (\bibinfo {year} {1978})\BibitemShut
  {NoStop}%
\bibitem [{\citenamefont {Bolognese}\ \emph {et~al.}(1979)\citenamefont
  {Bolognese}, \citenamefont {Engel}, \citenamefont {Guyonnet},\ and\
  \citenamefont {Riester}}]{Bolognese:1979gf}%
  \BibitemOpen
  \bibfield  {author} {\bibinfo {author} {\bibfnamefont {T.}~\bibnamefont
  {Bolognese}}, \bibinfo {author} {\bibfnamefont {J.~P.}\ \bibnamefont
  {Engel}}, \bibinfo {author} {\bibfnamefont {J.~L.}\ \bibnamefont
  {Guyonnet}},\ and\ \bibinfo {author} {\bibfnamefont {J.~L.}\ \bibnamefont
  {Riester}},\ }\bibfield  {title} {\bibinfo {title} {{Single pion production
  in antineutrino induced charged current interactions}},\ }\href
  {https://doi.org/10.1016/0370-2693(79)90361-7} {\bibfield  {journal}
  {\bibinfo  {journal} {Phys.\ Lett.}\ }\textbf {\bibinfo {volume} {81\,B}},\
  \bibinfo {pages} {393} (\bibinfo {year} {1979})}\BibitemShut {NoStop}%
\bibitem [{\citenamefont {Allen}\ \emph {et~al.}(1980)\citenamefont {Allen}
  \emph {et~al.}}]{Allen:1980ti}%
  \BibitemOpen
  \bibfield  {author} {\bibinfo {author} {\bibfnamefont {P.}~\bibnamefont
  {Allen}} \emph {et~al.} (\bibinfo {collaboration}
  {Aachen--Bonn--CERN--M$\ddot{\text{u}}$nchen--Oxford Collaboration}),\
  }\bibfield  {title} {\bibinfo {title} {{Single $\pi^+$ production in charged
  current neutrino-hydrogen interactions}},\ }\href
  {https://doi.org/10.1016/0550-3213(80)90450-2} {\bibfield  {journal}
  {\bibinfo  {journal} {Nucl.\ Phys.\ B}\ }\textbf {\bibinfo {volume} {176}},\
  \bibinfo {pages} {269} (\bibinfo {year} {1980})}\BibitemShut {NoStop}%
\bibitem [{\citenamefont {Allasia}\ \emph {et~al.}(1983)\citenamefont {Allasia}
  \emph {et~al.}}]{Allasia:1983qh}%
  \BibitemOpen
  \bibfield  {author} {\bibinfo {author} {\bibfnamefont {D.}~\bibnamefont
  {Allasia}} \emph {et~al.} (\bibinfo {collaboration}
  {Amsterdam--Bergen--Bologna--Padova--Pisa--Saclay--Torino Collaboration}),\
  }\bibfield  {title} {\bibinfo {title} {{Single pion production in charged
  current $\overline{\nu}D$ interactions at high energies}},\ }\href
  {https://doi.org/10.1007/BF01573212} {\bibfield  {journal} {\bibinfo
  {journal} {Z.\ Phys.\ C}\ }\textbf {\bibinfo {volume} {20}},\ \bibinfo
  {pages} {95} (\bibinfo {year} {1983})}\BibitemShut {NoStop}%
\bibitem [{\citenamefont {Barlag}(1984)}]{Barlag:1984uga}%
  \BibitemOpen
  \bibfield  {author} {\bibinfo {author} {\bibfnamefont {S.~J.~M.}\
  \bibnamefont {Barlag}},\ }\emph {\bibinfo {title} {{Quasielastic interactions
  and one pion production by neutrinos and antineutrinos on a deuterium
  target}}},\ \href@noop {} {Ph.D. thesis},\ \bibinfo  {school} {Amsterdam U.}
  (\bibinfo {year} {1984})\BibitemShut {NoStop}%
\bibitem [{\citenamefont {Allen}\ \emph {et~al.}(1986)\citenamefont {Allen}
  \emph {et~al.}}]{Allen:1985ti}%
  \BibitemOpen
  \bibfield  {author} {\bibinfo {author} {\bibfnamefont {P.}~\bibnamefont
  {Allen}} \emph {et~al.} (\bibinfo {collaboration}
  {Aachen--Birmingham--Bonn--CERN--London--Munich--Oxford Collaboration}),\
  }\bibfield  {title} {\bibinfo {title} {{A study of single-meson production in
  neutrino and antineutrino charged-current interactions on protons}},\ }\href
  {https://doi.org/10.1016/0550-3213(86)90480-3} {\bibfield  {journal}
  {\bibinfo  {journal} {Nucl.\ Phys.\ B}\ }\textbf {\bibinfo {volume} {264}},\
  \bibinfo {pages} {221} (\bibinfo {year} {1986})}\BibitemShut {NoStop}%
\bibitem [{\citenamefont {Jones}\ \emph {et~al.}(1989)\citenamefont {Jones}
  \emph {et~al.}}]{Jones:1989vt}%
  \BibitemOpen
  \bibfield  {author} {\bibinfo {author} {\bibfnamefont {G.~T.}\ \bibnamefont
  {Jones}} \emph {et~al.} (\bibinfo {collaboration} {Birmingham--CERN--Imperial
  College--M$\ddot{\text{u}}$nchen (MPI)--Oxford University College (WA21
  Collaboration)}),\ }\bibfield  {title} {\bibinfo {title} {{Experimental test
  of the PCAC hypothesis in the reactions $\nu_{\mu}p\to\mu^-p\pi^+$ and
  $\overline{\nu}_{\mu}p\to\mu^+p\pi^-$ in the $\Delta(1232)$ region}},\ }\href
  {https://doi.org/10.1007/BF01550930} {\bibfield  {journal} {\bibinfo
  {journal} {Z.\ Phys.\ C}\ }\textbf {\bibinfo {volume} {43}},\ \bibinfo
  {pages} {527} (\bibinfo {year} {1989})}\BibitemShut {NoStop}%
\bibitem [{\citenamefont {Allasia}\ \emph {et~al.}(1984)\citenamefont {Allasia}
  \emph {et~al.}}]{Allasia:1983dq}%
  \BibitemOpen
  \bibfield  {author} {\bibinfo {author} {\bibfnamefont {D.}~\bibnamefont
  {Allasia}} \emph {et~al.} (\bibinfo {collaboration}
  {Amsterdam--Bergen--Bologna--Padua--Pisa--Saclay--Torino Collaboration}),\
  }\bibfield  {title} {\bibinfo {title} {{Measurement of the $\nu_\mu$ and
  $\overline{\nu}_\mu$ -- nucleon charged current total cross-sections, and the
  ratio of $\nu_\mu$ neutron to $\nu_\mu$ proton charged current total
  cross-section}},\ }\bibfield  {booktitle} {\emph {\bibinfo {booktitle} {{11th
  International Symposium on Lepton and Photon Interactions at High Energies,
  Ithaca, New York, USA, August 4--9, 1983}}},\ }\href
  {https://doi.org/10.1016/0550-3213(84)90250-5} {\bibfield  {journal}
  {\bibinfo  {journal} {Nucl.\ Phys.\ B}\ }\textbf {\bibinfo {volume} {239}},\
  \bibinfo {pages} {301} (\bibinfo {year} {1984})}\BibitemShut {NoStop}%
\bibitem [{\citenamefont {Allasia}\ \emph {et~al.}(1990)\citenamefont {Allasia}
  \emph {et~al.}}]{Allasia:1990uy}%
  \BibitemOpen
  \bibfield  {author} {\bibinfo {author} {\bibfnamefont {D.}~\bibnamefont
  {Allasia}} \emph {et~al.},\ }\bibfield  {title} {\bibinfo {title}
  {{Investigation of exclusive channels in $\nu/\overline{\nu}$-deuteron
  charged current interactions}},\ }\href
  {https://doi.org/10.1016/0550-3213(90)90472-P} {\bibfield  {journal}
  {\bibinfo  {journal} {Nucl.\ Phys.\ B}\ }\textbf {\bibinfo {volume} {343}},\
  \bibinfo {pages} {285} (\bibinfo {year} {1990})}\BibitemShut {NoStop}%
\bibitem [{\citenamefont {Furuno}\ \emph {et~al.}(2002)\citenamefont {Furuno}
  \emph {et~al.}}]{Furuno:2003ng-proc}%
  \BibitemOpen
  \bibfield  {author} {\bibinfo {author} {\bibfnamefont {K.}~\bibnamefont
  {Furuno}} \emph {et~al.},\ }\bibfield  {title} {\bibinfo {title} {{BNL 7-foot
  bubble chamber experiment: Neutrino deuterium interactions}},\ }in\
  \href@noop {} {\emph {\bibinfo {booktitle} {{Proceedings of the 2nd
  International Workshop on Neutrino-Nucleus Interactions in the Few GeV Region
  (NuInt\,2002), Irvine, California, USA, December 12--15, 2002}}}},\ \bibinfo
  {editor} {edited by\ \bibinfo {editor} {\bibfnamefont {J.~G.}\ \bibnamefont
  {Morfin}}}\ (\bibinfo  {publisher} {North Holland Publishing Co.},\ \bibinfo
  {address} {Amsterdam, The Netherlands},\ \bibinfo {year} {2002})\BibitemShut
  {NoStop}%
\bibitem [{\citenamefont {Sakuda}\ and\ \citenamefont
  {Paschos}(2002)}]{Sakuda:2002-KK}%
  \BibitemOpen
  \bibfield  {author} {\bibinfo {author} {\bibfnamefont {M.}~\bibnamefont
  {Sakuda}}\ and\ \bibinfo {author} {\bibfnamefont {E.~F.}\ \bibnamefont
  {Paschos}},\ }\bibfield  {title} {\bibinfo {title} {{Single pion production
  in neutrino-nucleus interactions in the few GeV region}},\ }in\ \href@noop {}
  {\emph {\bibinfo {booktitle} {{Proceedings of the 2nd International Workshop
  on Neutrino-Nucleus Interactions in the Few GeV Region (NuInt\,2002), Irvine,
  California, USA, December 12--15, 2002}}}},\ \bibinfo {editor} {edited by\
  \bibinfo {editor} {\bibfnamefont {J.~G.}\ \bibnamefont {Morfin}}}\ (\bibinfo
  {publisher} {North Holland Publishing Co.},\ \bibinfo {address} {Amsterdam,
  The Netherlands},\ \bibinfo {year} {2002})\BibitemShut {NoStop}%
\bibitem [{\citenamefont {Sakuda}(2004)}]{Sakuda:2003-KK}%
  \BibitemOpen
  \bibfield  {author} {\bibinfo {author} {\bibfnamefont {M.}~\bibnamefont
  {Sakuda}},\ }\bibfield  {title} {\bibinfo {title} {{Study of neutrino-nucleus
  interactions for neutrino oscillations}},\ }in\ \href@noop {} {\emph
  {\bibinfo {booktitle} {{Proceedings of the 4th Workshop on Neutrino
  Oscillations and their Origin (NOON\,2003), Ishikawa Kousei Nenkin Kaikan,
  Kanazawa, Japan, February 10--14, 2003}}}},\ \bibinfo {editor} {edited by\
  \bibinfo {editor} {\bibfnamefont {Y.}~\bibnamefont {Suzuki}}, \bibinfo
  {editor} {\bibfnamefont {M.}~\bibnamefont {Nakahata}}, \bibinfo {editor}
  {\bibfnamefont {Y.}~\bibnamefont {Itow}}, \bibinfo {editor} {\bibfnamefont
  {M.}~\bibnamefont {Shiozawa}},\ and\ \bibinfo {editor} {\bibfnamefont
  {Y.}~\bibnamefont {Obayashi}}}\ (\bibinfo  {publisher} {World Scientific
  Publishing Co Pte Ltd},\ \bibinfo {address} {5 Toh Tuck Link, Singapore
  596224},\ \bibinfo {year} {2004}),\ pp.\ \bibinfo {pages}
  {253--260}\BibitemShut {NoStop}%
\bibitem [{\citenamefont {Rodrigues}\ \emph {et~al.}(2016)\citenamefont
  {Rodrigues}, \citenamefont {Wilkinson},\ and\ \citenamefont
  {McFarland}}]{Rodrigues:2016xjj}%
  \BibitemOpen
  \bibfield  {author} {\bibinfo {author} {\bibfnamefont {P.}~\bibnamefont
  {Rodrigues}}, \bibinfo {author} {\bibfnamefont {C.}~\bibnamefont
  {Wilkinson}},\ and\ \bibinfo {author} {\bibfnamefont {K.}~\bibnamefont
  {McFarland}},\ }\bibfield  {title} {\bibinfo {title} {{Constraining the GENIE
  model of neutrino-induced single pion production using reanalyzed bubble
  chamber data}},\ }\href {https://doi.org/10.1140/epjc/s10052-016-4314-3}
  {\bibfield  {journal} {\bibinfo  {journal} {Eur.\ Phys.\ J.\ C}\ }\textbf
  {\bibinfo {volume} {76}},\ \bibinfo {pages} {474} (\bibinfo {year} {2016})},\
  \Eprint {https://arxiv.org/abs/1601.01888} {arXiv:1601.01888 [hep-ex]}
  \BibitemShut {NoStop}%
\bibitem [{\citenamefont {Hawker}(2002)}]{Hawker:02}%
  \BibitemOpen
  \bibfield  {author} {\bibinfo {author} {\bibfnamefont {E.~A.}\ \bibnamefont
  {Hawker}},\ }\bibfield  {title} {\bibinfo {title} {{Single pion production in
  low energy neutrino-carbon}},\ }in\ \href@noop {} {\emph {\bibinfo
  {booktitle} {{Proceedings of the 2nd International Workshop on
  Neutrino-Nucleus Interactions in the Few GeV Region (NuInt\,2002), Irvine,
  California, USA, December 12--15, 2002}}}},\ \bibinfo {editor} {edited by\
  \bibinfo {editor} {\bibfnamefont {J.~G.}\ \bibnamefont {Morfin}}}\ (\bibinfo
  {publisher} {North Holland Publishing Co.},\ \bibinfo {address} {Amsterdam,
  The Netherlands},\ \bibinfo {year} {2002})\BibitemShut {NoStop}%
\bibitem [{\citenamefont {Kakorin}\ \emph {et~al.}(2020)\citenamefont
  {Kakorin}, \citenamefont {Kuzmin},\ and\ \citenamefont
  {Naumov}}]{Kakorin:2020atz}%
  \BibitemOpen
  \bibfield  {author} {\bibinfo {author} {\bibfnamefont {I.~D.}\ \bibnamefont
  {Kakorin}}, \bibinfo {author} {\bibfnamefont {K.~S.}\ \bibnamefont
  {Kuzmin}},\ and\ \bibinfo {author} {\bibfnamefont {V.~A.}\ \bibnamefont
  {Naumov}},\ }\bibfield  {title} {\bibinfo {title} {{A unified empirical model
  for quasielastic interactions of neutrino and antineutrino with nuclei}},\
  }\href {https://doi.org/10.3390/sym12081285} {\bibfield  {journal} {\bibinfo
  {journal} {Phys.\ Part.\ Nucl.\ Lett.}\ }\textbf {\bibinfo {volume} {17}},\
  \bibinfo {pages} {265} (\bibinfo {year} {2020})}\BibitemShut {NoStop}%
\bibitem [{\citenamefont {Barish}\ \emph
  {et~al.}(1977{\natexlab{b}})\citenamefont {Barish} \emph
  {et~al.}}]{Barish:1977ny}%
  \BibitemOpen
  \bibfield  {author} {\bibinfo {author} {\bibfnamefont {B.~C.}\ \bibnamefont
  {Barish}} \emph {et~al.},\ }\bibfield  {title} {\bibinfo {title}
  {{Measurements of $\nu_{\mu}N$ and $\overline{\nu}_{\mu}N$ charged current
  total cross sections}},\ }\href {https://doi.org/10.1103/PhysRevLett.39.1595}
  {\bibfield  {journal} {\bibinfo  {journal} {Phys.\ Rev.\ Lett.}\ }\textbf
  {\bibinfo {volume} {39}},\ \bibinfo {pages} {1595} (\bibinfo {year}
  {1977}{\natexlab{b}})}\BibitemShut {NoStop}%
\bibitem [{\citenamefont {Bosetti}\ \emph {et~al.}(1977)\citenamefont {Bosetti}
  \emph {et~al.}}]{Bosetti:1977nd}%
  \BibitemOpen
  \bibfield  {author} {\bibinfo {author} {\bibfnamefont {P.~C.}\ \bibnamefont
  {Bosetti}} \emph {et~al.} (\bibinfo {collaboration}
  {Aachen--Bonn--CERN--London--Oxford-Saclay}),\ }\bibfield  {title} {\bibinfo
  {title} {{Total cross sections for charged current neutrino and antineutrino
  interactions in BEBC in the energy range 20 -- 200 GeV}},\ }\href
  {https://doi.org/10.1016/0370-2693(77)90537-8} {\bibfield  {journal}
  {\bibinfo  {journal} {Phys.\ Lett.}\ }\textbf {\bibinfo {volume} {70B}},\
  \bibinfo {pages} {273} (\bibinfo {year} {1977})}\BibitemShut {NoStop}%
\bibitem [{\citenamefont {Gl$\ddot{\text{u}}$ck}\ \emph
  {et~al.}(1998)\citenamefont {Gl$\ddot{\text{u}}$ck}, \citenamefont {Reya},\
  and\ \citenamefont {Vogt}}]{Gluck:1998xa}%
  \BibitemOpen
  \bibfield  {author} {\bibinfo {author} {\bibfnamefont {M.}~\bibnamefont
  {Gl$\ddot{\text{u}}$ck}}, \bibinfo {author} {\bibfnamefont {E.}~\bibnamefont
  {Reya}},\ and\ \bibinfo {author} {\bibfnamefont {A.}~\bibnamefont {Vogt}},\
  }\bibfield  {title} {\bibinfo {title} {{Dynamical parton distributions
  revisited}},\ }\href {https://doi.org/10.1007/s100529800978,
  10.1007/s100520050289} {\bibfield  {journal} {\bibinfo  {journal} {Eur.\
  Phys.\ J.\ C}\ }\textbf {\bibinfo {volume} {5}},\ \bibinfo {pages} {461}
  (\bibinfo {year} {1998})},\ \Eprint {https://arxiv.org/abs/hep-ph/9806404}
  {hep-ph/9806404} \BibitemShut {NoStop}%
\bibitem [{\citenamefont {Grabosch}\ \emph {et~al.}(1989)\citenamefont
  {Grabosch} \emph {et~al.}}]{Grabosch:1988gw}%
  \BibitemOpen
  \bibfield  {author} {\bibinfo {author} {\bibfnamefont {H.~J.}\ \bibnamefont
  {Grabosch}} \emph {et~al.} (\bibinfo {collaboration} {SKAT Collaboration}),\
  }\bibfield  {title} {\bibinfo {title} {{Cross section measurements of single
  pion production in charged current neutrino and antineutrino interactions}},\
  }\href {https://doi.org/10.1007/BF01564697} {\bibfield  {journal} {\bibinfo
  {journal} {Z.\ Phys.\ C}\ }\textbf {\bibinfo {volume} {41}},\ \bibinfo
  {pages} {527} (\bibinfo {year} {1989})}\BibitemShut {NoStop}%
\bibitem [{\citenamefont {Nakamura}\ \emph {et~al.}(2019)\citenamefont
  {Nakamura}, \citenamefont {Kamano},\ and\ \citenamefont
  {Sato}}]{Nakamura:2018ntd}%
  \BibitemOpen
  \bibfield  {author} {\bibinfo {author} {\bibfnamefont {S.~X.}\ \bibnamefont
  {Nakamura}}, \bibinfo {author} {\bibfnamefont {H.}~\bibnamefont {Kamano}},\
  and\ \bibinfo {author} {\bibfnamefont {T.}~\bibnamefont {Sato}},\ }\bibfield
  {title} {\bibinfo {title} {{Impact of final state interactions on
  neutrino--nucleon pion production cross sections extracted from
  neutrino--deuteron reaction data}},\ }\href
  {https://doi.org/10.1103/PhysRevD.99.031301} {\bibfield  {journal} {\bibinfo
  {journal} {Phys.\ Rev.\ D}\ }\textbf {\bibinfo {volume} {99}},\ \bibinfo
  {pages} {031301} (\bibinfo {year} {2019})},\ \Eprint
  {https://arxiv.org/abs/1812.00144} {arXiv:1812.00144 [hep-ph]} \BibitemShut
  {NoStop}%
\end{thebibliography}
  \end{document}